\documentclass[twocolumn,aps,prb,showpacs,superscriptaddress,amsfonts,amssymb,amsmath]{revtex4}
\usepackage{amsmath}
\usepackage{graphicx}
\begin{document}
\title{Possible existence of a band of extended states induced by inter-Landau-band mixing in quantum Hall system}
\author{Gang Xiong}
\affiliation{Physics Department, Beijing Normal University,
Beijing 100875, P. R. China}
\author{Shi-Dong Wang}
\affiliation{Physics Department, Hong Kong University of Science
and Technology, Clear Water Bay, Hong Kong SAR, China}
\author{Qian Niu}
\affiliation{Physics Department, The University of Texas at
Austin, Austin, Texas 78712-1081}
\affiliation{International Center for Quantum Structures,
Institute of Physics, Chinese Academy of Sciences , Beijing
100080, P. R. China}
\author{Yupeng Wang}
\affiliation{International Center for Quantum Structures,
Institute of Physics, Chinese Academy of Sciences , Beijing
100080, P. R. China}
\author{X. C. Xie}
\affiliation{Physics Department, Oklahoma State University ,
Stillwater, OK 74078}
\affiliation{International Center for Quantum Structures,
Institute of Physics, Chinese Academy of Sciences , Beijing
100080, P. R. China}
\author{De-Cheng Tian}
\affiliation{Physics Department, Wuhan
University, Wuhan 430072, P. R. China}
\author{X. R. Wang}
\affiliation{Physics Department, Hong Kong
 University of Science and Technology, Clear Water Bay,
Hong Kong SAR, China}

\date{Draft on \today}

\begin{abstract}
The mixing of states with {\it opposite chirality} in quantum Hall
system is shown to have delocalization effect. It is possible that
extended states may form bands because of this mixing, as is shown
through a numerical calculation on a two-channel network model.
Based on this result, a new phase diagram with a narrow {\it
metallic} phase separating two adjacent QH phases and/or
separating a QH phase from the insulating phase is proposed. The
data from recent non-scaling experiments is reanalyzed and shown
that they seem to be {\it consistent} with the new phase diagram.
However, due to finite-size effects, further study on large system
size is still needed to conclude whether there are extended state
bands in thermodynamic limit.
\end{abstract}

 \maketitle

\section{Introduction}
According to the scaling theory of localization\cite {abrahams},
all electrons in a disordered two-dimensional system are localized
in the absence of a magnetic field. In the presence of a strong
magnetic field $B$, a series of disorder-broadened Landau bands
(LBs) will appear, and an extended state resides at the center of
each band while states at other energies are
localized\cite{pruisken}. The integrally quantized Hall plateaus
(IQHP) are observed when the Fermi level lies in localized states,
with the value of the Hall conductance, $\sigma_{xy}=ne^2/h$,
related to the number of occupied extended states($n$). Many
previous studies\cite{wei,pruisken2,kivelson,dzliu,wang,yang,
galstyan,sheng1,sheng2,haldane,jiang,shahar1,tkwang,glozman,kravchenko,
song,schaijk,xrw,hilke,baba,shash} have been focused on so-called
plateau transitions. The issue there is how the Hall conductance
jumps from one quantized value to another when the Fermi level
crosses an extended state. There are two competing proposals. One
is the global phase diagram\cite{kivelson} based on the levitation
of extended states conjectured by Khmelnitskii\cite{khme} and
Laughlin\cite{laughlin}. A crucial prediction of this phase
diagram is that an integer quantum Hall effect (IQHE) state $n$ in
general can only go into another IQHE states $n\pm 1$, and that a
transition into an insulating state is allowed only from the $n=1$
state. The other is so-called direct transition phase
diagram\cite{dzliu} in which transitions from any IQHE state to
the insulating phase are allowed when the disorder is increased at
fixed $B$. So far, most experiments\cite{kravchenko,song} are
consistent with the direct transition phase diagram although the
early experiments were interpreted in terms of the global phase
diagram.

One important yet overlooked issue regarding IQHE is the {\it
nature} of both plateau-plateau and plateau-insulator transitions.
In all existing theoretical studies, these transitions are assumed
to be continuous quantum phase transitions. This assumption is
mainly due to the early scaling experiments\cite{wei,pruisken2}.
The fingerprint of a continuous phase transition is scaling laws
around the transition point. In the case of IQHE, it means
algebraic divergence of longitudinal resistance slope and
algebraic shrinkage of longitudinal resistance peak width in
temperature $T$ at the transition point. However, some experiments
show that longitudinal resistance slope remains finite\cite{hilke}
and resistance peak width remains nonzero\cite{shash,baba} when
extrapolated to zero temperature. This implies a {\it non-scaling}
behavior around a transition point, contradicting the expectation
of continuous quantum phase transitions suggested by the theories.
Thus the nature of these transitions should be re-examined.

The samples used in the non-scaling experiments\cite{hilke} are
relatively dirty, and strong disorders should lead to a strong
inter-Landau-band mixing. In a recent Letter\cite {xiong}, we
showed that the single extended state at each LB center broadens
into a narrow band of extended states when the interband mixing of
{\it opposite chirality} is taken into account. A narrow metallic
phase exists between two adjacent IQHE phases and between an IQHE
phase and the insulating phase. A plateau-plateau or
plateau-insulator transition corresponds to two consecutive
quantum phase transitions instead of one as suggested by existing
theories. This possibility is usually overlooked in previous
studies where each longitudinal conductance peak is related to a
single extended state\cite{glozman}. In this paper we shall
present the detailed description of this study.

The paper is organized as follows. The semiclassical network model
for two coupled LBs is illustrated in Sec. II. It is shown that
mixing of localized states of {\it opposite chirality} tends to
delocalize a state while mixing of states of the same chirality
does not. The similarities and differences between the case we
consider and models for double layer systems and spin-degenerate
systems are also discussed. Our approach, level-statistics
technique, is described in Sec. III. Sec. IV has four subsections
containing our numerical results, conclusions and discussions.
Sec. IV A provides our main numerical results and conclusions. In
Sec. IV B, possible finite size effects and a theoretical
understanding for non-scaling behaviors based on standard scaling
theory\cite{pruisken3,huckstein} are discussed. Sec. IV C gives
further numerical results which directly show a narrow band of
extended states in several cases, and a new phase diagram of
quantum Hall systems is proposed. To further support our results,
in Sec. IV D we reanalyze the original data from the non-scaling
experiments\cite{hilke} and show that the non-scaling behaviors in
each IQHP-insulator transition can be attributed to two quantum
phase transition points. Recent
experiments\cite{hohls1,hohls2,hohls3,ponomarenko,li}, in which
scaling behaviors are observed, are also discussed and shown to be
not inconsistent with early non-scaling experiments and our
numerical results. The conclusions of this paper are summarized in
Sec. V.

\section{The semiclassical model including
inter-Landau-band mixing}
According to the semiclassical
theory\cite{chalker}, the motion of an electron in a strong
magnetic field and in a smooth random potential can be decomposed
into a rapid cyclotron motion and a slow drifting motion of its
guiding center. The kinetic energy of the cyclotron motion is
quantized by $E_n=(n+1/2)\hbar\omega_c$, where $\omega_c$ is the
cyclotron frequency and $n$ the LB index. The trajectory of the
drifting motion of the guiding center is thus along an
equipotential contour of value $V_0=E-E_n$, where $E$ is the total
energy of the electron. The local drifting velocity
$\vec{v}(\vec{r})$ is determined by (in SI unit)
\begin{equation}
\vec{v}(\vec{r})=\bigtriangledown V
(\vec{r})\times\vec{B}/(eB^2)
\label{drift}
\end{equation}
where $\bigtriangledown V(\vec{r})$ is the local potential
gradient. The equipotential contour consists of many closed loops.
Neglect quantum tunnelling effects, each loop corresponds to
trajectory of one eigenstate. The motion of electrons are thus
confined around these loops with deviations typically order of the
cyclotron radius $l_c=\sqrt{\hbar/(eB)}$.
\begin{figure}[ht]
\begin{center}
 \includegraphics[height=6cm, width=6cm]{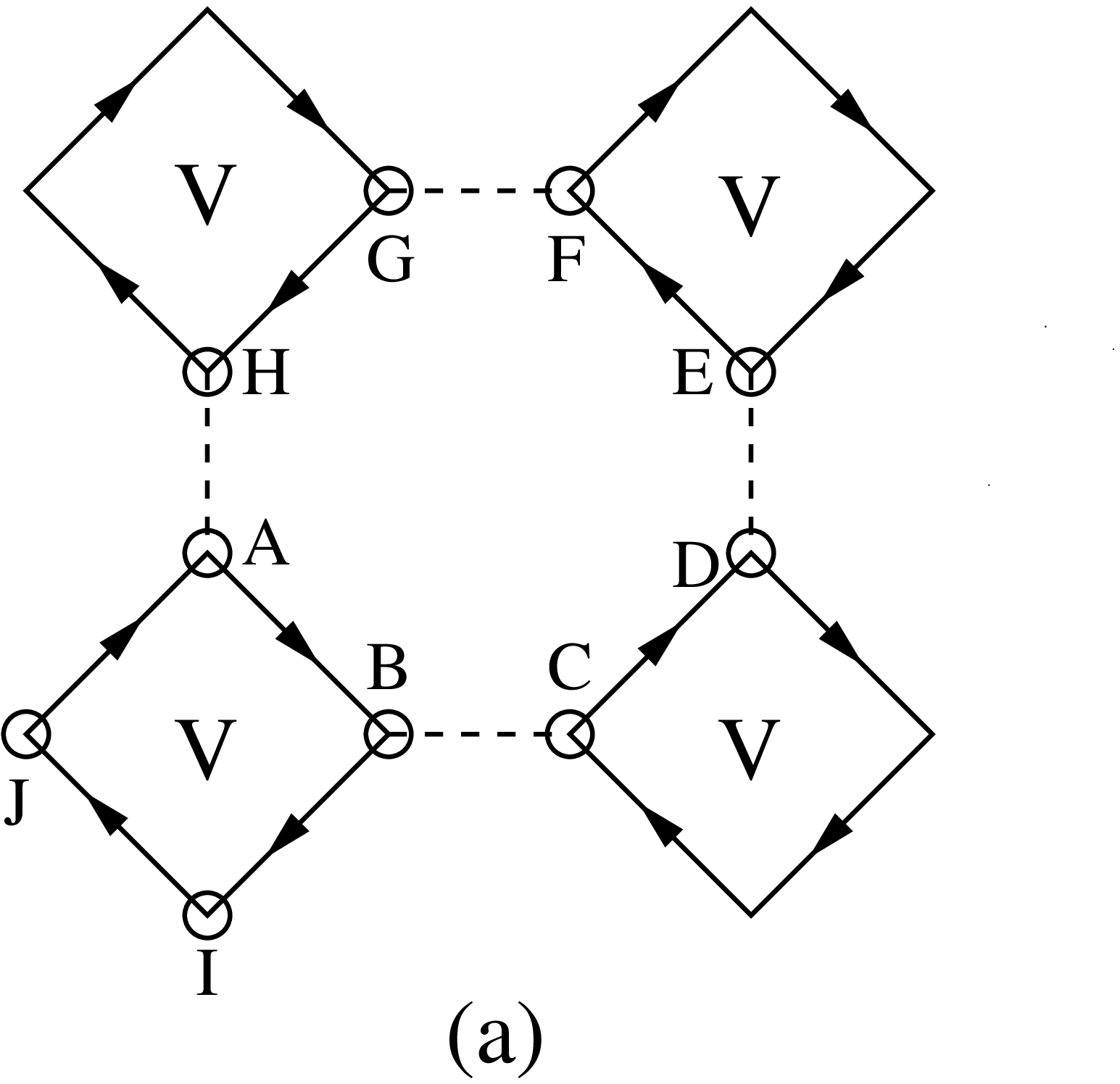}
 \includegraphics[height=5.5cm, width=4cm]{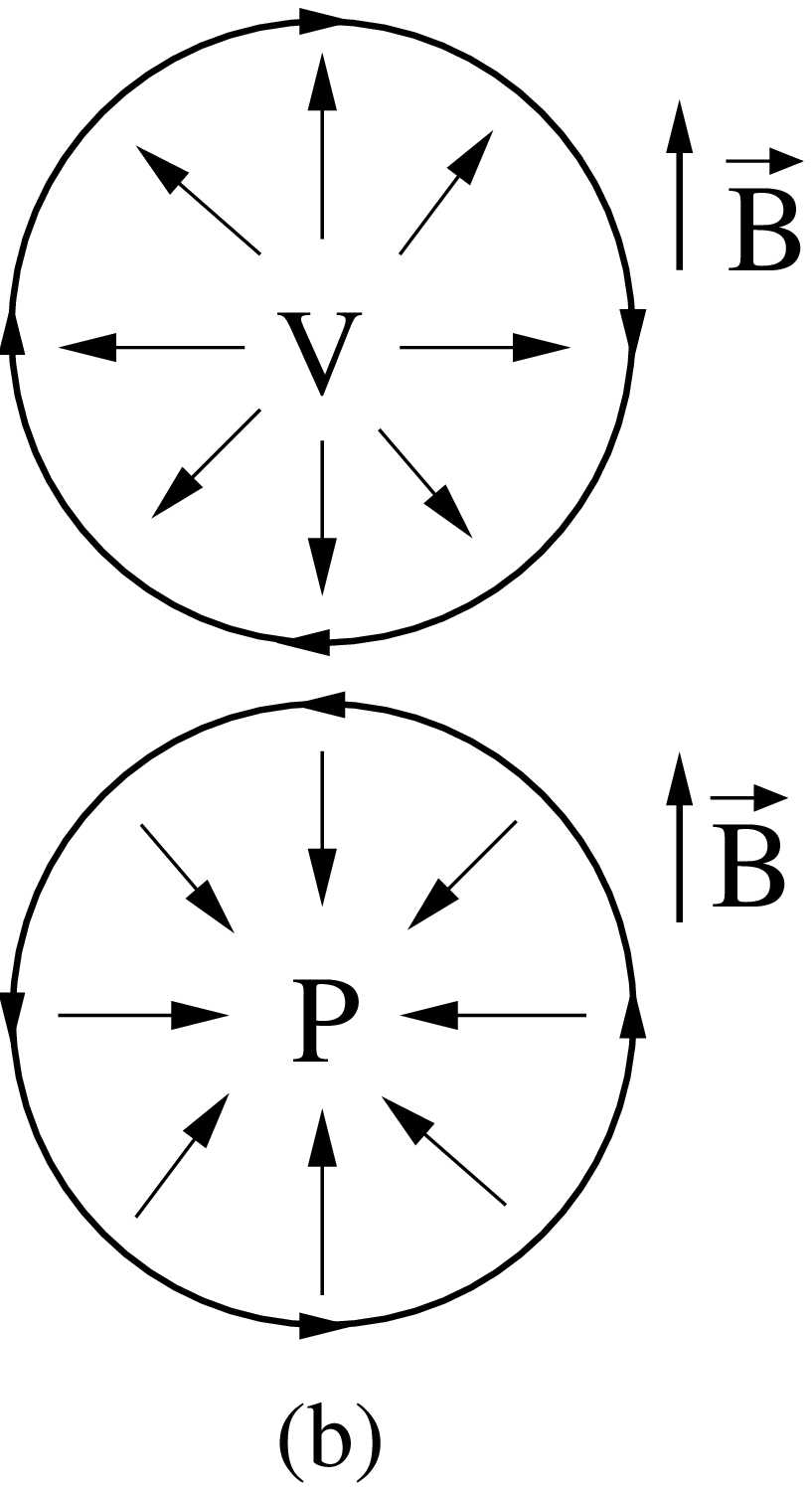}
\end{center}
\caption{(a) Four neighboring loops in a one-band model for the
case of $V_0<0$. Dashed lines denote quantum tunnellings. The
arrows indicate the drifting direction. (b) Loops localized around
a valley and a peak, respectively. The arrows inside a loop show
the directions of local potential gradient around the peak or
valley. The arrows on a loop indicate the drifting direction.}
\label{1channel}
\end{figure}

To illustrate this semiclassical picture, let us think of the
smooth random potential as a landscape of many peaks and valleys
distributed randomly in the plane. Imagine that the landscape is
filled with water up to a height of value $V_0$.
The equipotential contour of value $V_0$ is the boundary of
land and water. According to the percolation
theory\cite{stauff}, the percolation threshold of a
two-dimensional (2D) continuum model is $p_c=1/2$, where $p_c$ is
the occupation probability of the medium (the land or the water).
For simplicity, we suppose that the distribution of the random
potential is symmetric around zero. By symmetry the percolation
point of both the land and the water is at $V_0=0$ in this case.
When $V_0<0$, the occupation probability of land is above $1/2$.
Thus the land percolates and the water forms isolated lakes. These
lakes are around valleys and their boundaries correspond to
trajectories of localized states. In the case of $V_0>0$, the
water forms a percolating sea and the land becomes isolated
islands around potential peaks. The boundary of each island is an
electronic state. In short, semiclassical electronic states in a
QH system are equipotential loops. These loops are localized
around potential peaks for $V_0>0$ and around potential valleys
for $V_0<0$. The drifting direction of each loop is {\it
unidirectional}. This means that they are chiral states. From Eq.
\ref{drift} one can see that states around peaks have {\it
opposite chirality} from states around valleys because the
directions of the local potential gradient around a peak is
opposite to that around a valley. If one views the plane from the
direction opposite to the magnetic field, the drifting is
clockwise around valleys and counter-clockwise around peaks, as
shown in Fig.\ref{1channel}. Right at $V_0=0$ both the land and
the water percolate, and the intersection between them is the
trajectory of an extended state. It means that there is only one
extended state at $V_0=0$ for each LB. As $V_0$ approaches zero
from both sides, the localization length $\xi$ of the system tends
to diverge as
\begin{equation}
    \xi\propto |V_0|^{-\nu}
\label{ll}
\end{equation}
where the critical exponent $\nu=4/3$ according to the classical
percolation theory. Quantum effects are ignored in the above
semiclassical argument. When two spatially separated loops on the
same equipotential contour come close at saddle points of the
random potential, quantum tunnellings should be considered. An
example in the case of $V_0<0$ is shown in Fig.\ref{1channel}(a).
In the absence of interband mixing, numerical calculations have
suggested that there is still only one extended state in each LB
while the value of the critical exponent $\nu$ is modified to be
around $7/3$\cite{galstyan}.
\begin{figure}[ht]
\begin{center}
 \includegraphics[height=7cm,width=8cm]{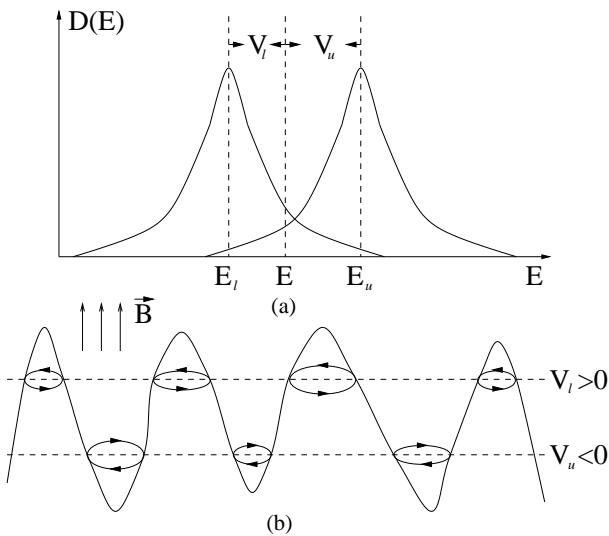}
\end{center}
\caption{(a) Two adjacent Landau bands in the case when the
disorder broadened band width is comparable with the Landau gap.
$D(E)$ is the density of states. $E_{\it u}$ and $E_{\it l}$
denote the centers of the two bands. (b) Schematic plot of the two
sets of equipotential loops on the two-dimensional random
potential for electronic states of energy $E$ shown in (a). It is
the projection of a three-dimensional landscape on a 2D vertical
plane. The solid curve represents the schematic plot of the 2D
random potential. Two dashed horizontal lines indicate two
constant potential planes $V(\vec{r})\equiv V_{0}$, one for the
lower band with $V_0= V_{\it l}=E-E_{\it u}>0$ and the other for
the upper band with $V_0=V_{\it u}=E-E_{\it l}<0$. The ellipses
denote the loops where the two constant potential planes intersect
with the 2D random potential. Arrows on the loops show drifting
directions. $\vec{B}$ is the magnetic field.} \label{twoband}
\end{figure}

\begin{figure}[h]
\begin{center}
\includegraphics[height=7.5cm, width=7.5cm]{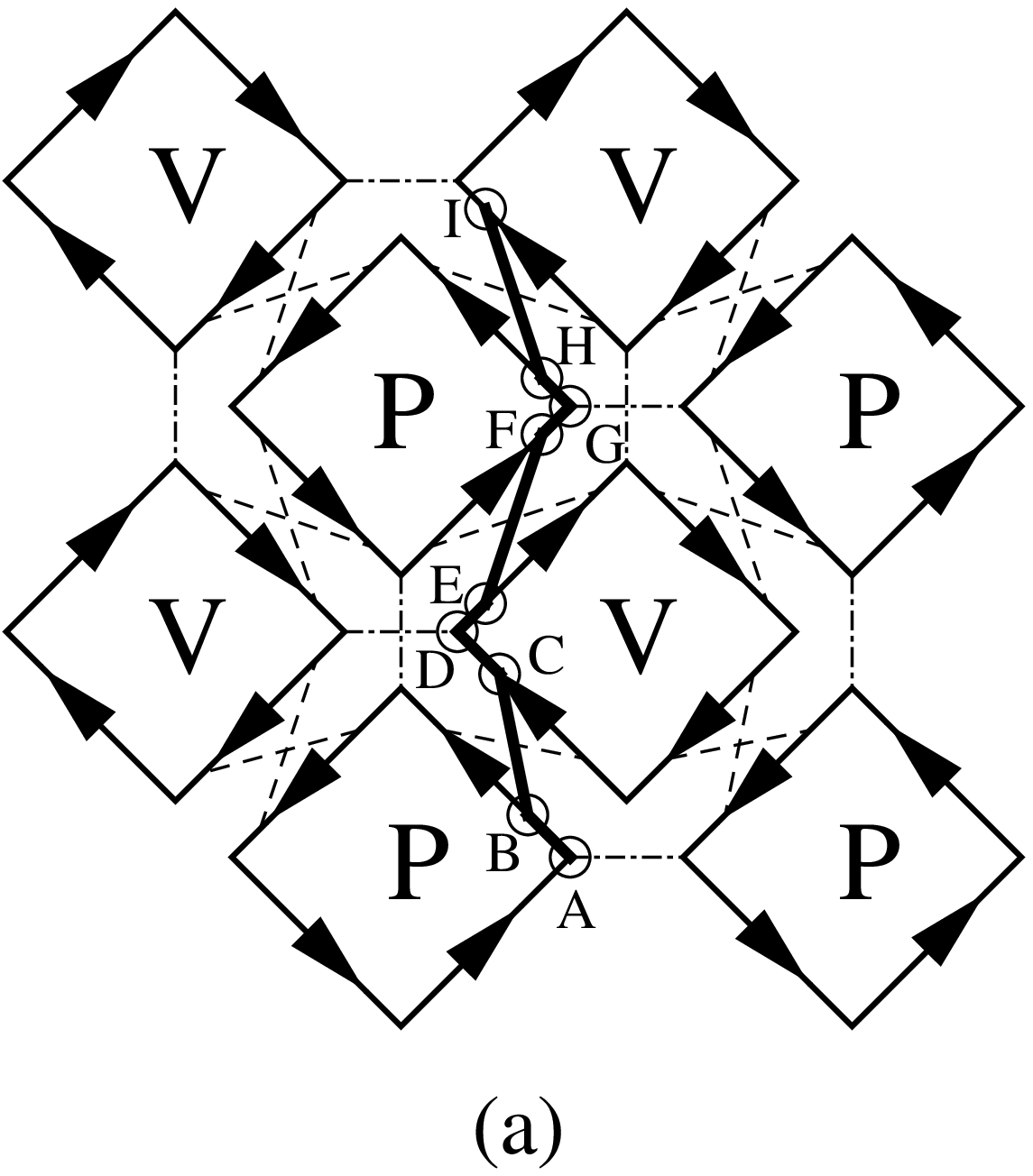}
\includegraphics[height=7.5cm, width=7.5cm]{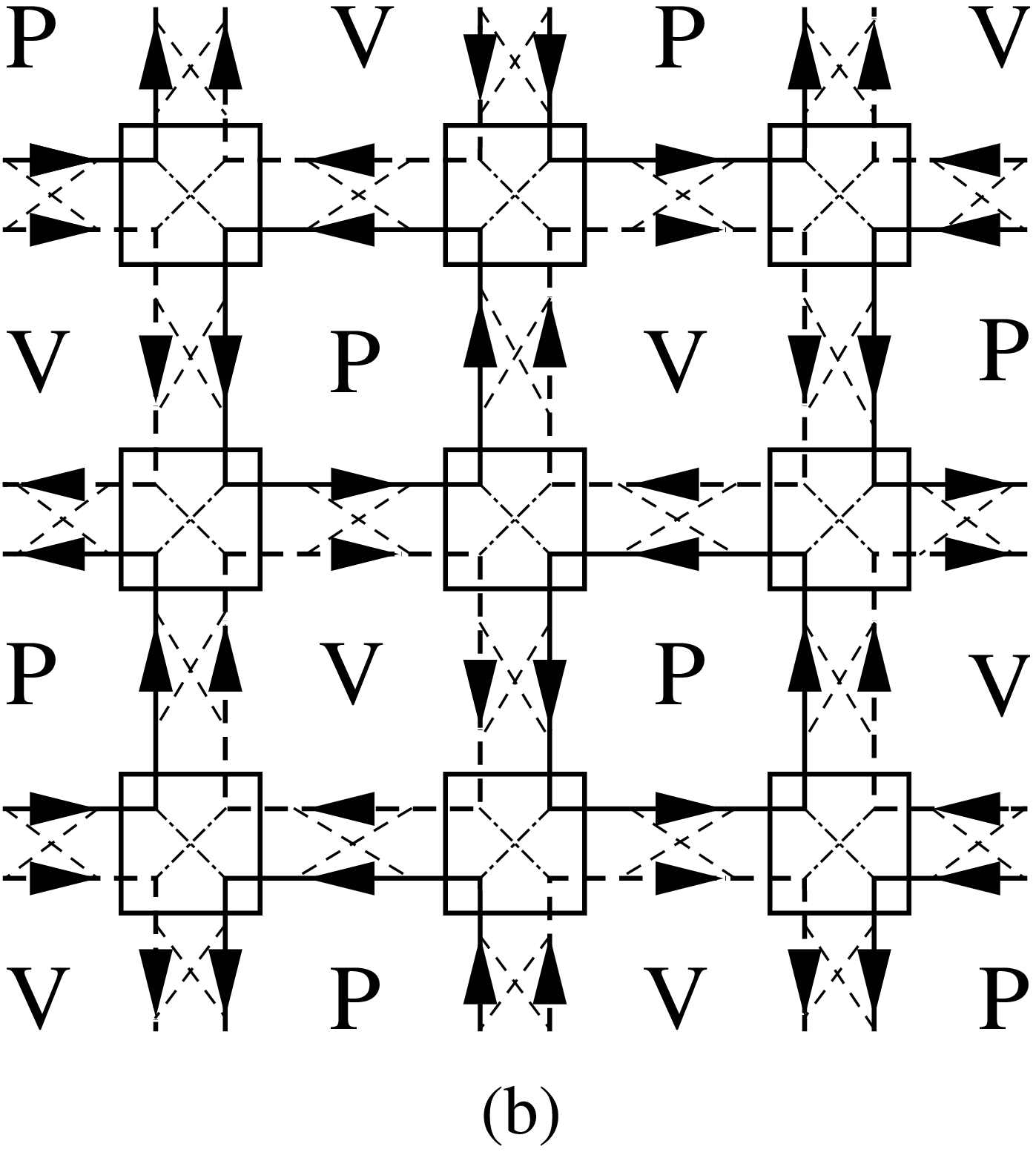}
\end{center}
\caption{(a) Topological plot of the trajectories of the drifting
motion of guiding centers (rhombus). The drifting motion around a
potential peak (valley) is denoted by P (V), and their directions
are indicated by the arrows. Dashed lines stand for interband
mixing, and dash-dotted lines for tunnelling at saddle points. The
thick line (A to I) describes the trajectory of an electron due to
a strong interband mixing. (b) The equivalent two-channel network
model of (a). Solid and dashed lines on each link denote two
channels from two LBs. Squares stand for saddle points. P, V and
arrows have the same meaning as those in (a).} \label{network}
\end{figure}

In the case when the width of the LBs is comparable with the
spacing between adjacent LBs (the Landau gap), and
inter-Landau-band mixing should no longer be ignored. In order to
investigate the consequences of inter-Landau-band mixing, we shall
consider a simple system of two adjacent LBs. Since we are
interested in interband-mixing of opposite chirality, we consider
those states with energy between the lower and the upper bands
whose centers are at $E_l$ and $E_u$, respectively, as shown in
Fig.\ref{twoband}(a). Thus, equipotential loops are $V_l=E-E_l>0$
and $V_u=E-E_u<0$ for the lower and the upper LBs, respectively.
Using the semiclassical theory described in the previous
paragraphs, states from the upper band should move along
equipotential loops around potential valleys while those from the
lower band around potential peaks as shown in
Fig.\ref{twoband}(b). The loops for the upper band drift in
clockwise direction, and those for the lower band in the
counter-clockwise direction. These two sets of loops are thus
spatially separated and have {\it opposite chirality}. If we
assume that peaks and valleys of random potential form two coupled
square lattices, the loops can be arranged as shown in Fig.
\ref{network}(a), where {\it P} and {\it V} denote peaks and
valleys, respectively. In the absence of interband mixing, the
model is reduced to two decoupled single-band models and all
electronic states between the two LBs are localized. If we
introduce interband mixing, the localized loops may become less
localized. To see this, let us consider an extreme case with no
tunnellings at saddle points, but with such a strong interband
mixing that an electron will move from a loop around a valley to
its neighboring loop around a peak and vice versa, as shown by
$B\to C$ in Fig.\ref{network}(a). Follow the trajectory of an
electron starting at A, it will be $A\to B\to C\to D\to E \cdot
\cdot \cdot$. The electron is no longer confined on a closed loop,
but delocalized!

Before going on, we would like to make a short discussion on the
relation between large disorder magnitude and weak magnetic field.
The above semiclassical picture for the network model is valid
only when the typical fluctuation length of random potential,
denoted by $L_F$, is much larger than the magnetic length
$l_c=\sqrt{\hbar/(eB)}$. Let us take $L_F$ as a large fixed value.
Then, the effect of inter-band mixing is controlled mainly by two
parameters. One is the ratio of disorder magnitude $W$ and Landau
gap $\hbar\omega_c$, which is proportional to $W/B$ and determines
whether inter-band mixing needs to be considered. The other is the
ratio of the magnetic length $l_c$ and the typical distance
between an upper-band loop and its neighboring lower-band loop
$L_F\frac{\hbar\omega_c}{W}$, which is proportional to $W/B^{3/2}$
and determines the typical strength of inter-band mixing in the
network model. Therefore, the cases of strong disorder $W$ and
weak magnetic field $B$ are, to some extent, equivalent in the
network model since both of them lead to large inter-band mixing.
Of cause, values of local potential gradients are also important,
since they determine the detailed distribution of inter-band
mixing in the model.

In the one-band model, an electron can also hop from one loop to
its neighboring loops by quantum tunnellings. At a first glance,
this effect seems similar to that of interband mixing. However,
they are fundamentally different. In the one-band model,
electronic states for a given $V_0$ have {\it the same chirality}.
Thus the drifting direction of an electron will be inverted when
it tunnels into neighboring loops. This means that a strong
tunnelling in a one-band model will induce an effective
backward-scattering that tends to also localize electrons. We can
understand this by considering a small part of the one-band model
as shown in Fig.\ref{1channel}(a) where all loops are moving in
clockwise direction. Without tunnelling, the trajectory of an
electron starting from point A is $A\to B\to I \to J\to A$, a
clockwise closed loop. With strong tunnellings, the trajectory
will tend to be $A\to B\to C\to D \cdot \cdot \cdot \to H \to A$,
a counter-clockwise closed loop. Thus, the tunnellings between
loops of the same chirality cannot delocalize states.

Two-channel CC model can also be used to simulate both
spin-degenerated and bilayer systems. The model for
spin-degenerate system is the same as we use. The only difference
is that loops for spin-up and spin-down states present at the same
positions in real space, and the chiralities of loops of spin-up
and spin-down states are the same. While the main difference
between the model for bilayer system and ours is that the real
space distributions of random potentials in the two layers are
different.

It is worthwhile to explain why we consider only those states
between two LB centers. For states outside this range, both sets
of loops are localized around either valleys or peaks. This means
that interband mixing mainly occurs between two loops localized
around the same position, and this mixing will not delocalize a
state. In fact, as explained in the previous paragraph, the mixing
of the same chirality does not help delocalizing an electron. This
is why we shall consider mixing between spatially separated states
with opposite chirality. Of course, it does not mean that the
mixing of the same chirality has no effect at all. As it was found
in some previous works\cite{wang}, this kind of mixing may shift
an extended state from its LB center. Level shifting due to mixing
between states of the same chirality may distort the shape of the
phase diagram, but should not alter its topology. The emergence of
the bands of extended states is exclusively due to the mixing
between states of opposite chirality.

\begin{figure}[ht]
\begin{center}
 \includegraphics[height=5.0cm,width=8cm]{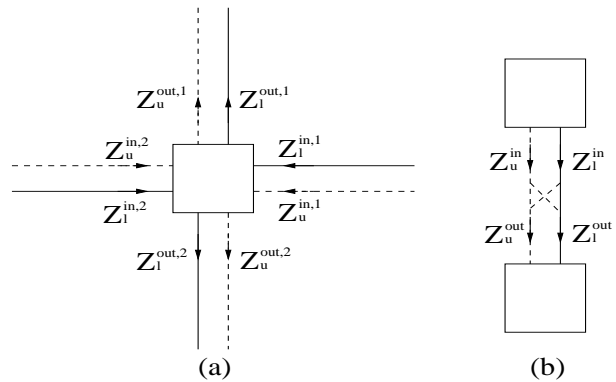}
\end{center}
\caption{(a) A node with four incoming and four outgoing channels.
$Z_{u(l)}^{in,i}$ is the wavefunction amplitude of the i$^{th}$
incoming wave from the upper (lower) LB. $Z_{u(l)}^{out,i}$ is
that of outgoing wavefunction amplitude. (b) A link with two
channels. $Z_{u(l)}^{in (out)}$ is the incoming (outgoing)
wavefunction amplitude of the upper (lower) LB.} \label{nodelink}
\end{figure}

Now, we describe our two-channel network model in detail. Assume
that tunnellings of two neighboring localized states (loops) of
the same band occur around saddle points, and interband mixing
takes place only on the links, Fig.\ref{network}(a) is
topologically equivalent to the model shown in
Fig.\ref{network}(b). Fig.\ref{network}(b) is the schematic
illustration of our {\it two-channel Chalker-Coddington} network
model. It is similar to the model studied in previous
publications\cite{lee, kagalovsky}. There are two channels on each
link. One, denoted by a solid line, is from the lower LB around a
potential peak. The other (dashed line) is from the upper LB
moving around a potential valley. The arrows indicate the drifting
direction of the two sets of states. At each node, the tunnelling
between two neighboring states of the same LB occurs. As shown in
Fig.\ref{nodelink}(a), let $Z_{u(l)}^{in,1}$ and $Z_{u(l)}^{in,2}$
be the incoming wave amplitudes of states 1 and 2 from upper
(lower) LB, respectively, and $Z_{u(l)}^{out,1}$ and
$Z_{u(l)}^{out,2}$ be the outgoing wave amplitudes of the two
states. The tunnelling is described by an SO(4) matrix
\begin{equation}
\label{smatrix}
    \left (
    \begin{array}{l}
    Z_u^{out,1}\\
    Z_u^{out,2}\\
    Z_l^{out,1}\\
    Z_l^{out,2}
    \end{array}
    \right )
    =
    \left (
    \begin{array}{llll}
    s_u^R & s_u^L & 0 & 0 \\
    -s_u^L & s_u^R & 0 & 0 \\
    0 & 0 & s_l^R & s_l^L \\
    0 & 0 & -s_l^L & s_l^R
    \end{array}
    \right )
    \left (
    \begin{array}{l}
    Z_u^{in,1}\\Z_u^{in,2}\\Z_l^{in,1}\\Z_l^{in,2}
    \end{array}
    \right ) ,
\end{equation}
where the subscripts $u$ and $l$ denote the upper and the lower
bands, respectively. The elements $s_{u(l)}^L$ and $s_{u(l)}^R$
are tunnelling coefficients of an incoming wave-function in the
upper (lower) band being scattered into outgoing channels at its
left-hand and right-hand sides, respectively. $s_{u(l)}^R$ and
$s_{u(l)}^L$ are related to each other as $s_{u(l)}^R=\sqrt
{1-(s_{u(l)}^L)^2}$ due to the orthogonality of the matrix. Under
quadratic potential barrier approximation, ---i.e.,
$V(x,y)=-Ux^2+Uy^2+V_c$ around a saddle point, where $U$ is a
constant describing the strength of potential fluctuation and
$V_c$ is the potential barrier at the point,--- one can show that
the left-hand scattering amplitude is given by\cite{fertig1}
\begin{equation}
    s_{u(l)}^L=[1+\exp(-\pi\epsilon_{u(l)})]^{-1/2},
\end{equation}
where $\epsilon_{u(l)}=[E+V_c-(n_{u(l)}+1/2)E_2]/E_1$ with $E$
being electron energy, $E_1=\frac{\hbar\omega_c}{2\sqrt{2}}
\sqrt{K-1}$, $E_2=\frac{\hbar\omega_c}{\sqrt{2}}\sqrt{K+1}$,
and $K=\sqrt{\frac{64U^2}{m^2\omega_c^4}+1}$. The energies of
the cyclotron motion in the two bands are $(n_u+1/2)E_2$ and
$(n_l+1/2)E_2$, respectively, where $n_{u(l)}$ are the band
indices and $\Delta n=n_u-n_l=1$. The dimensionless ratio
$E_r=E_2/E_1=2\sqrt {1+\frac{2}{K-1}}$ approaches $2$ from above
as $U$ or the inverse of $\omega_c$ increases\cite{fertig1},
i.e., strong disorder regime or a weak magnetic field.
In our calculations, we choose it to be 2.2 since this is
the regime we are interested in. For convenience, we choose
$E_2$ as the energy unit and the cyclotron energy of the lower
band as the reference point. The energy regime between the two
band centers is thus $E\in[0,1]$.

Inter-band mixing between two channels on a link as shown in
Fig.\ref{nodelink}(b) is described by a U(2) matrix
\begin{equation}
\label{mmatrix1}
    \left (
    \begin{array}{l} Z^{out}_l\\Z^{out}_u \end{array}
    \right )=M
    \left (
    \begin{array}{l} Z^{in}_l\\Z^{in}_u \end{array}
    \right ),
\end{equation}
\begin{equation}
\label{mmatrix2}
    M=\left (
    \begin{array}{ll} e^{i\phi_1} & 0 \\ 0 & e^{i\phi_2}
    \end{array}
    \right )
    \left (
    \begin{array}{ll} \cos\theta & \sin\theta \\ -
    \sin\theta & \cos\theta \end{array}
    \right )
    \left (
    \begin{array}{ll} e^{i\phi_3} & 0 \\ 0 & e^{i\phi_4}
    \end{array}
    \right ),
\end{equation}
where $\sin\theta$ describes the interband mixing. $\phi_i(i=1\sim
4)$ are random Aharonov-Bohm phases accumulated along propagation
paths. In our calculations, we shall assume that they are
uniformly distributed in $[0,2\pi]$\cite{chalker}. In the
following discussion, a parameter $J$, defined as $\sqrt{J/(1+
J)}=\sin\theta$, is used to characterize the mixing strength.
$J$ will take the same value for all links in our calculations.
We hope that this simplification will not affect the physics.

\section{The application of level-statistics technique
on the network model}

Electron localization length is often obtained from the transfer
matrix method. For a two-dimensional system, however, it is well
known that this quantity alone does not provide conclusive answers
to questions related to the metal-insulator transition
(MIT)\cite{xie}. On the other hand, level-statistics analysis has
been used in studying MIT\cite{kramer,schweitzer}.
Level-statistics analysis is based on random matrix theory
(RMT)\cite{mehta}. The basic idea is that the localization
property of an electronic state can be determined by the
statistical distribution function $P(s)$ of the spacing $s$ of two
neighboring levels. For localized states, the distribution is
Poissonian $P_{PE}(s)=exp(-s)$, called `Poissonian ensemble (PE)'.
In the case of extended states, the nearest neighbor level spacing
distribution has the following form\cite{mehta}
\begin{equation}
 P(s)=C_1(\beta)s^{\beta}exp[-C_2(\beta)s^2]
\end{equation}
where $C_1(\beta)$ and $C_2(\beta)$ are normalization factors
determined by $\int P(s)ds=1$ and $\int sP(s)ds=1$. The parameter
$\beta$ is determined by the dynamical symmetry of the system. The
case of $\beta=1$ is for systems with time-reversal symmetry and
an integer total angular momentum and is referred as `Gaussian
orthogonal ensemble'. Systems with time-reversal symmetry and a
half-integer total angular momentum belong to the case of
$\beta=4$, called `Gaussian symplectic ensemble'. For systems
without time-reversal symmetry $\beta=2$, and it is called
`Gaussian unitary ensemble (GUE)'. A fundamental difference
between level statistics property of extended states and localized
states is that $P(s)$ at $s=0$ is zero for extended states and one
for localized states. The physical reason of this difference is
the so-called `level repulsion' of extended states. Two extended
states with the same `bare energy' will overlap in real
space(since they are extended in real space) and form two new
extended states with different energies. While localized states
can have the same energy staying in different regions of real
space. This approach is appropriate for the network model of
quantum Hall systems because localized states in the model are
loops at different regions of real space while extended states in
the model are formed by quantum percolation of such localized
loops.

We shall follow the approach proposed by Klesse and
Metzler\cite{klesse}. A quantum state of a network model can be
expressed by a vector whose components are electronic
wave-function amplitudes on the links. In our case, the vector can
be written as $\Phi=(\{Z_u^i,Z_l^i\})$, where $Z_u^i$ and $Z_l^i$
are the electron wave-function amplitudes of the upper band ({\it
u}) and the lower band ({\it l}) on the $i$-th link, respectively.
As shown by Fertig\cite{fertig2}, the network model can be
described by an {\it evolution operator} $\hat{U}(E)$, an
$E$-dependent matrix determined by the scattering properties of
nodes and links in the model. (As an example, the evolution
operator of a two-channel network of size ${\it L}=2$ with
periodic boundaries on both directions is constructed explicitly
in the Appendix.) In general, the eigenvalue equation of the
evolution operator is
\begin{equation}
\hat{U}(E)\Phi_{\alpha}(E)=
e^{i\omega_{\alpha}(E)}\Phi_{\alpha}(E),
\end{equation}
where $\alpha$ is the eigenstate index of $\hat U$. The true
eigenenergies $\{E_n\}$ of the system are those energies at which
$\omega_\alpha(E)$ is an integer multiple of $2\pi$. It has been
shown by Klesse and Metzler\cite{klesse} that the set of {\it
quasi-energies} $\{\omega_{\alpha}(E)\}$ corresponds to the
excitation spectrum of the stationary state with energy $E$.
Therefore, the statistics property of the set of {\it
quasi-energies} $\{\omega_{\alpha}(E)\}$ at $E$ is the same as
that of true eigenenergies $\{E_n\}$ around $E$, and the
localization property of an electronic state with an energy $E$
can be obtained by the quasi-energies. The advantage of this
approach is that all the quasi-energies can be used in the
analysis so that better statistics can be obtained.

Chalker and Coddington\cite{chalker} showed numerically that an
open boundary condition along one direction creates extended edge
states along the other direction. In order to get rid of the edge
states, we employ a periodic boundary along both directions in our
calculation. For a two-channel network model of ${\it L}\times
{\it L}$ nodes with periodic boundaries along both directions,
there are $4{\it L}^2$ components in $\Phi$. $\hat{U}$ is thus a
$(4{\it L}^2)\times (4{\it L}^{2})$ matrix. However, there is a
special property of the network model\cite{metzler}: the nodes
scatter electrons only from vertical channels into horizontal
channels and {\it vice versa}. If one separates $\Phi$ into the
set of wavefunction amplitudes on the horizontal links $\Phi_H$
and the set of wavefunction amplitudes on the vertical links
$\Phi_V$, the evolution equation in one time step can be written
in the following form
\begin{equation}
\label{evolution}
    \left (
    \begin{array}{l}
    \Phi_H(t+1) \\ \Phi_V(t+1)
    \end{array}
    \right )
    =
    \left (
    \begin{array}{ll}
    \hat{0} & \hat{U}_{V\to H} \\
        \hat{U}_{H\to V} & \hat{0}
    \end{array}
    \right )
    \left (
    \begin{array}{l}
    \Phi_H(t) \\ \Phi_V(t)
    \end{array}
    \right ),
\end{equation}
where $\hat{0}$ is the $(2{\it L}^2)\times (2{\it L}^{2})$ zero
matrix. $\hat{U}_{V\to H}$ describes how wavefunction on vertical
links evolves into that on the horizontal  links. Similarly,
$\hat{U} _{H\to V}$ describes that from horizontal to vertical
links. For the detail derivation, we refer readers to the example
shown in the Appendix. The evolution equation in two time steps is
given as
\begin{eqnarray}
\Phi_H(t+2)=
\hat{U}_{V\to H}\hat{U}_{H\to V}\Phi_H(t) \\
\Phi_V(t+2)=
\hat{U}_{H\to V}\hat{U}_{V\to H}\Phi_V(t)
\end{eqnarray}
Therefore, the evolution matrix in two time steps is
block-diagonal and the two blocks have essentially the same
statistical property. We thus need only consider one of them.

We study the model for {\it L}=8, 12, 16,  20 and 24. The
calculation procedure is as follows. Take a realization of the
random phases, construct the evolution matrix and obtain the
quasi-energies $\{\omega_{i}\}$. Put them in descending order and
calculate the level spacings $s_i=(\omega_i-\omega_{i-1})
/\delta$, where $\delta$ is the average of ${s_i}$. Repeat this
procedure for sufficient times so that more than $5\times10^4$
level spacings are collected for a given $E$ and $J$. The
level-spacing distribution function $P(s)$ is thus obtained
numerically.

\section{Numerical results and discussions}
\subsection{Analysis of the level-spacing spectrum}

\begin{figure}[ht]
\begin{center}
 \includegraphics[width=8cm,height=4.9cm]{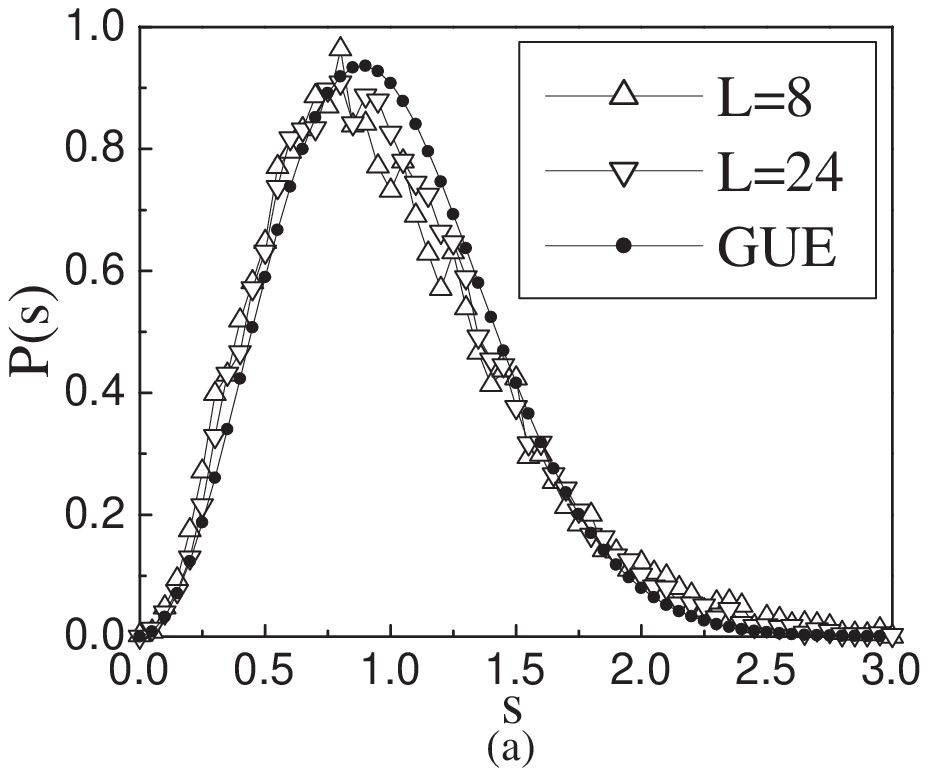}
\end{center}
\begin{center}
 \includegraphics[width=8cm,height=4.9cm]{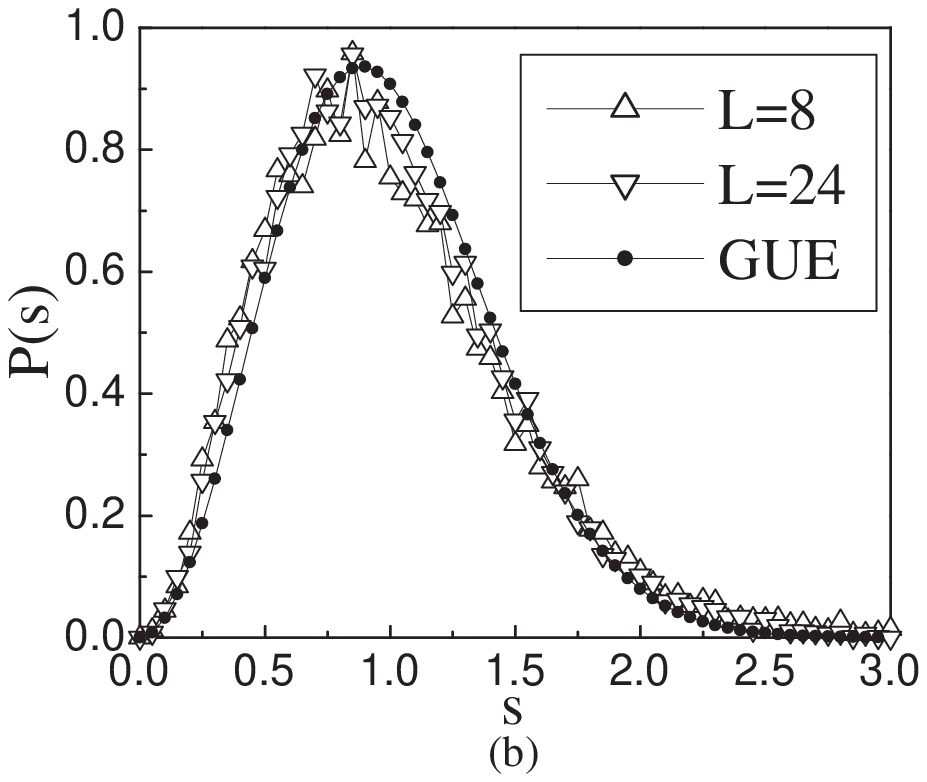}
\end{center}
\begin{center}
  \includegraphics[width=8cm,height=4.9cm]{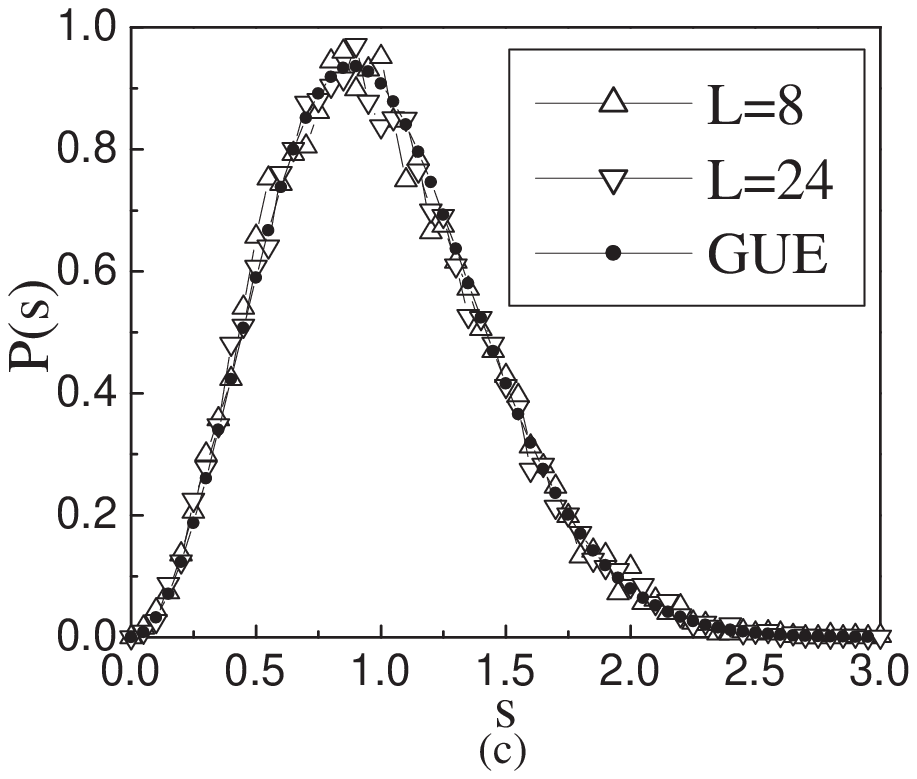}
\end{center}
\caption{$P(s)$ vs. $s$ for ${\it L}=8,24$ and $P_{GUE}(s)$. (a)
$E=0$ and $J=0.1$; (b) $E=0.02$ and $J=0.1$; (c) $E=0.5$ and
$J=1.5$.} \label{ps_global1}
\end{figure}

\begin{figure}[ht]
\begin{center}
 \includegraphics[width=8cm,height=4.9cm]{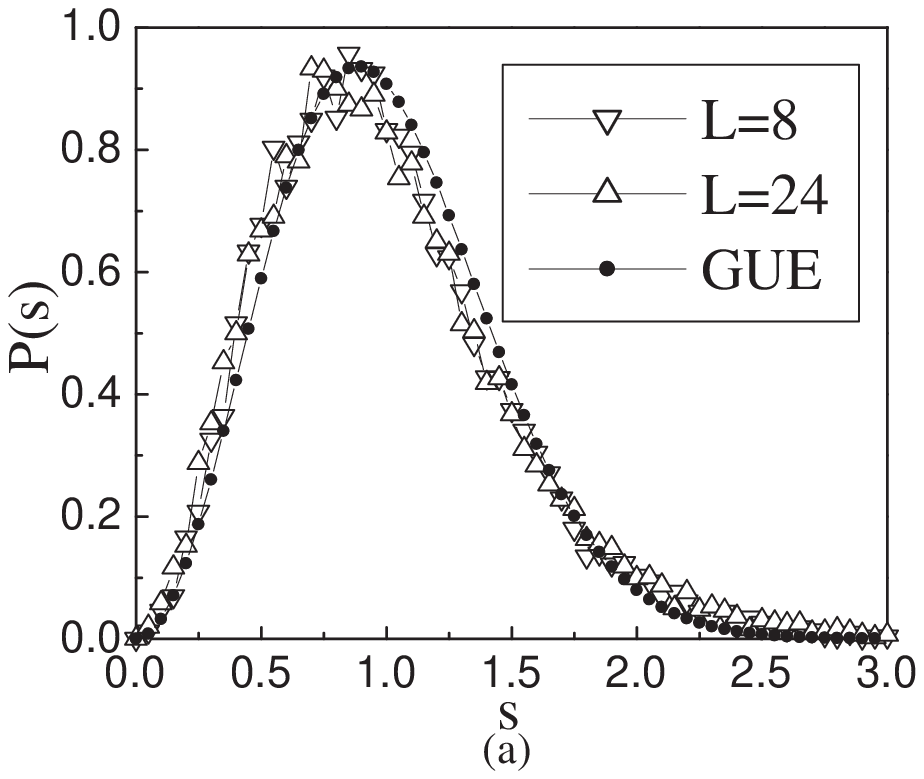}
\end{center}
\begin{center}
 \includegraphics[width=8cm,height=4.9cm]{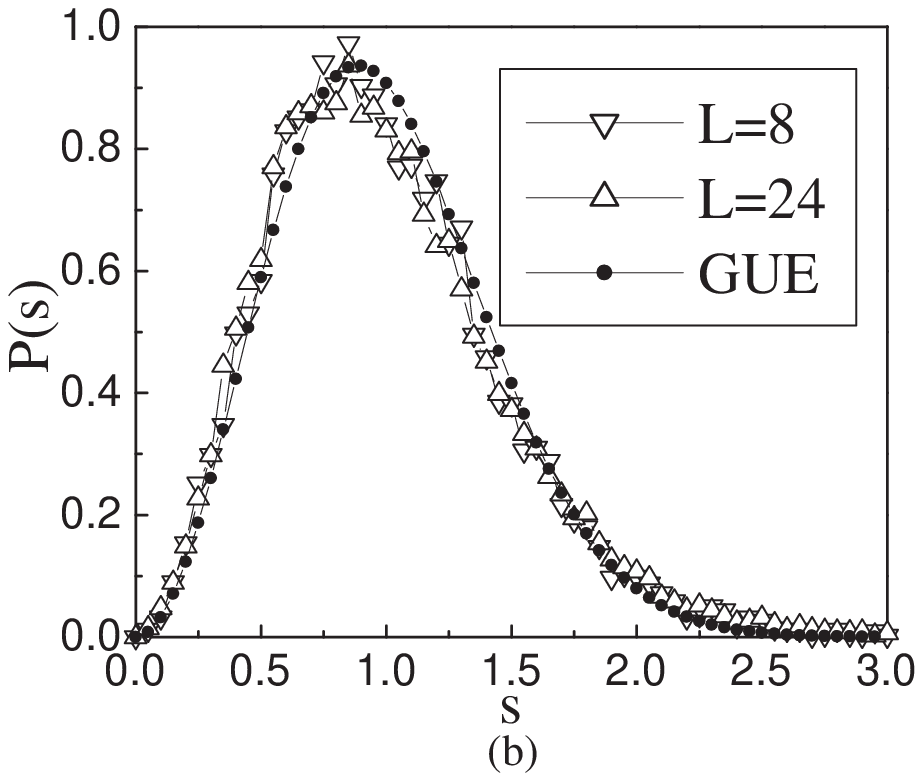}
\end{center}
\begin{center}
 \includegraphics[width=8cm,height=4.9cm]{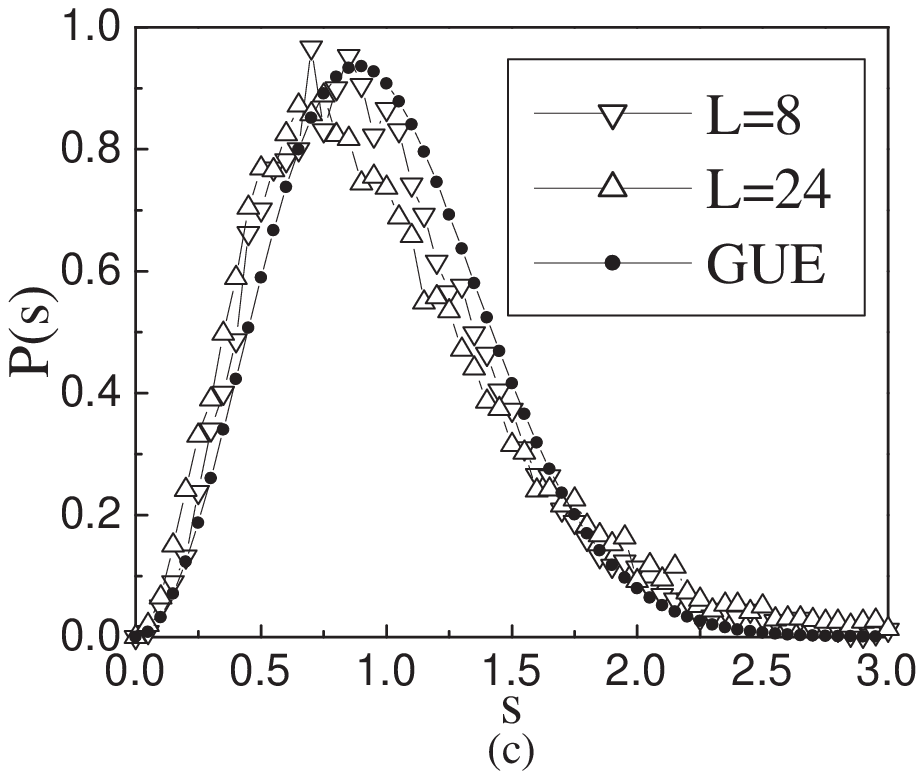}
\end{center}
\caption{$P(s)$ vs. $s$ for ${\it L}=8,24$ and $P_{GUE}(s)$. (a)
$E=0$ and $J=0.7$; (b) $E=0.02$ and $J=0.7$; (c) $E=0.5$ and
$J=0.5$.} \label{ps_global2}
\end{figure}

We shall analyze the numerical results of the level-spacing
distribution function $P(s)$ in order to provide evidences for the
existence of extended state bands in our model. Due to the
chirality nature of the drifting motion, time-reversal symmetry is
absent from our semiclassical network model. Then, according to
the RMT\cite{mehta}, $P(s)$ should be the GUE distribution,
$P_{GUE}(s)=32\pi^{-2}s^2exp(-4s^2/\pi)$, for extended states, and
the PE distribution $P_{PE}(s)$ for localized states. Since the
overall shape of the GUE distribution is quite different from that
of the PE distribution, one may use $P(s)$ to distinguish an
extended state from a localized state. Fig.\ref{ps_global1} is
$P_{GUE}(s)$ and $P(s)$ for $(E=0,J=0.1)$ (a), $(E=0.02, J=0.1)$
(b), and $(E=0.5,J=1.5)$ (c) with ${\it L}=8, 12,16,20,24$.
Fig.\ref{ps_global2} is $P(s)$ for $(E=0.0,J=0.7)$ (a),
$(E=0.02,J=0.7)$ (b) and $(E=0.5,J=0.5)$ (c).
\begin{figure}[ht]
\begin{center}
  \includegraphics[width=8cm,height=4.9cm]{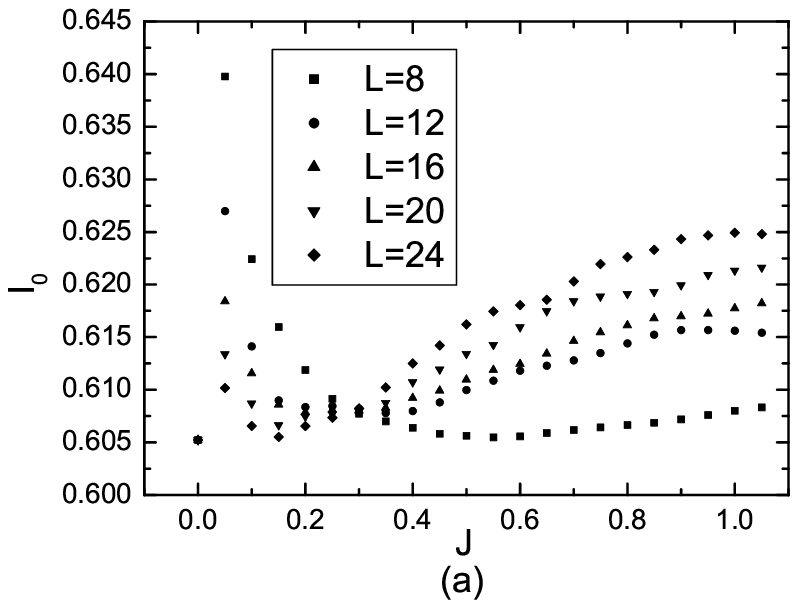}
\end{center}
\begin{center}
 \includegraphics[width=8cm,height=4.9cm]{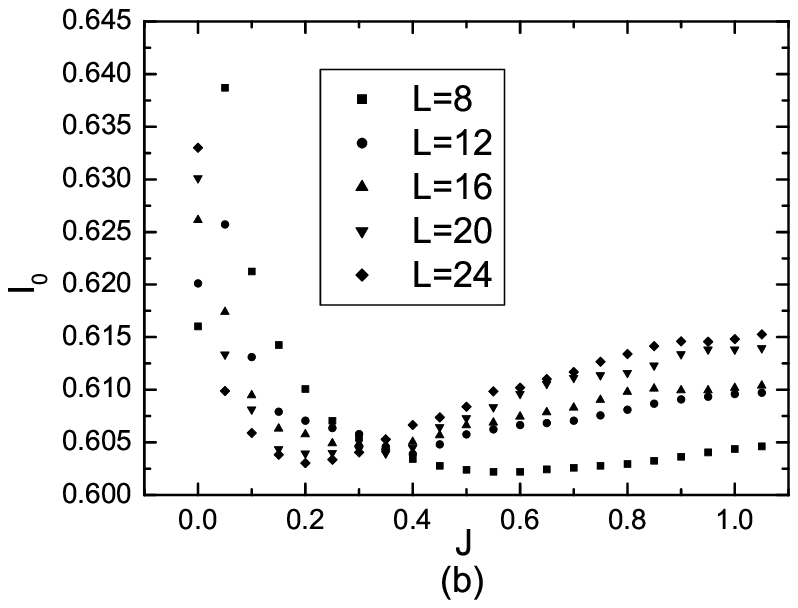}
\end{center}
\begin{center}
 \includegraphics[width=8cm,height=4.9cm]{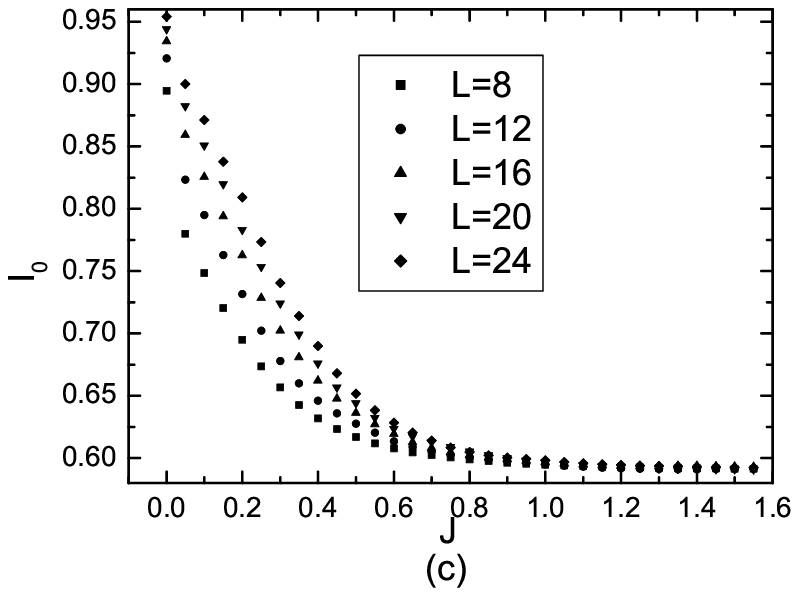}
\end{center}
\caption{$I_0$ vs. $J$ for ${\it L}=8, 12, 16, 20, 24$, (a) $E=0$;
(b) $E=0.02$; (c) $E=0.5$.} \label{data}
\end{figure}
\begin{figure}[ht]
\begin{center}
 \includegraphics[width=8cm,height=4.9cm]{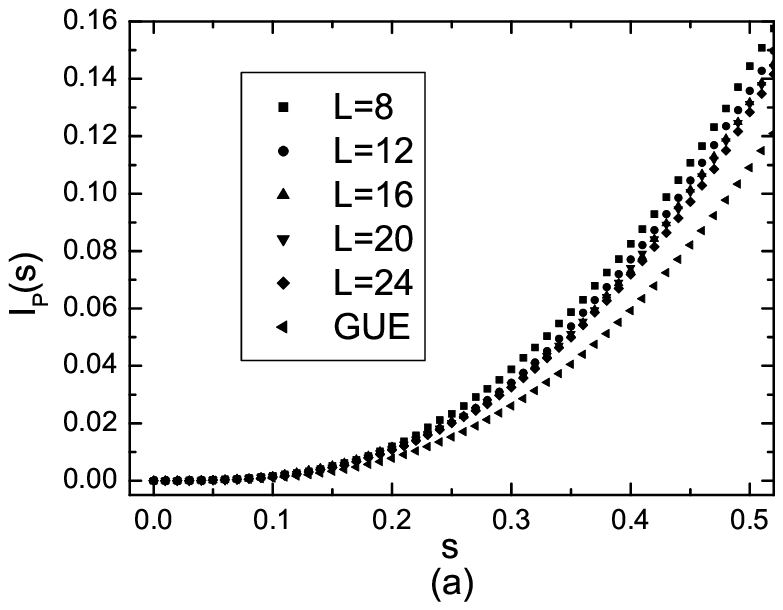}
\end{center}
\begin{center}
 \includegraphics[width=8cm,height=4.9cm]{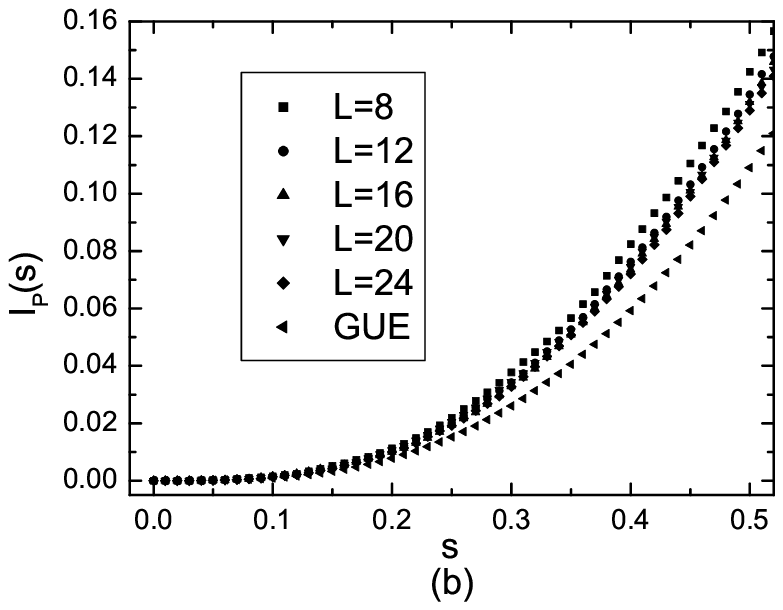}
\end{center}
\begin{center}
 \includegraphics[width=8cm,height=4.9cm]{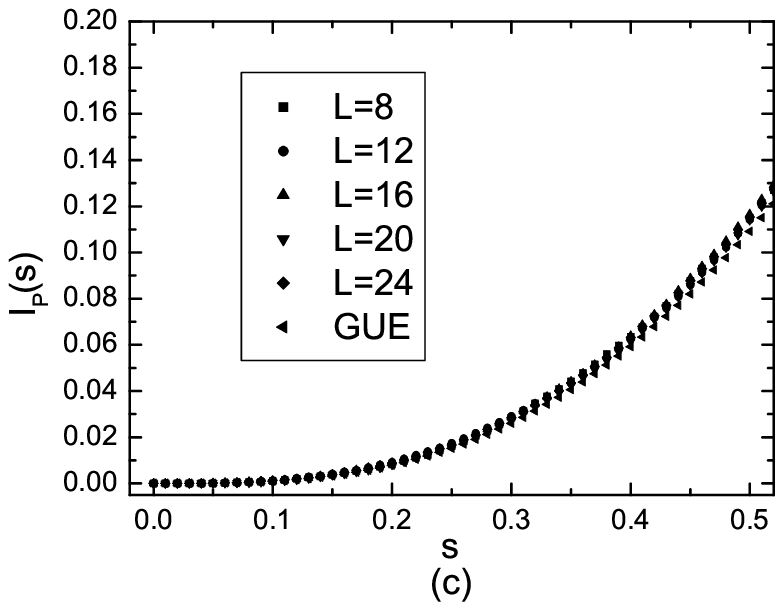}
\end{center}
\caption{$I_P(s)$ vs. $s$ for ${\it L}=8, 12, 16, 20, 24$ and that
for $P_{GUE}(s)$. (a) $E=0$ and $J=0.1$; (b) $E=0.02$ and $J=0.1$;
(c) $E=0.5$ and $J=1.5$.} \label{ps_small1}
\end{figure}
\begin{figure}[ht]
\begin{center}
 \includegraphics[width=8cm,height=4.9cm]{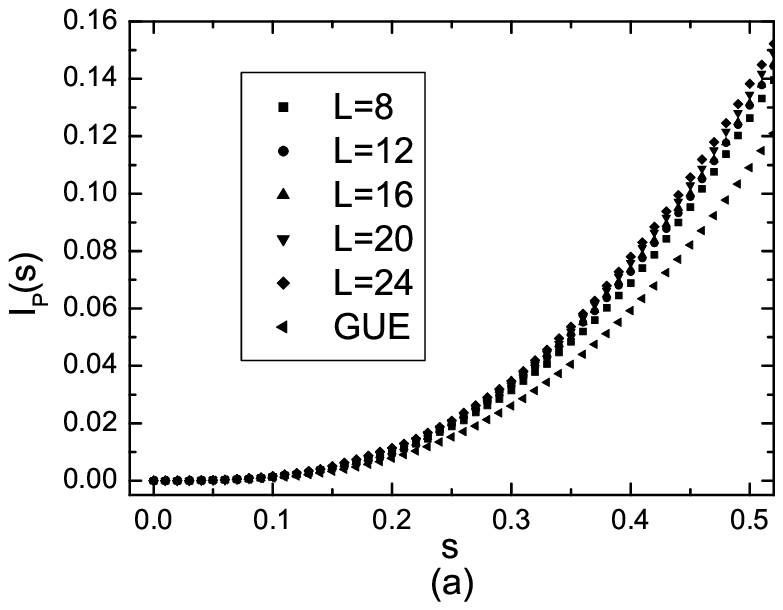}
\end{center}
\begin{center}
 \includegraphics[width=8cm,height=4.9cm]{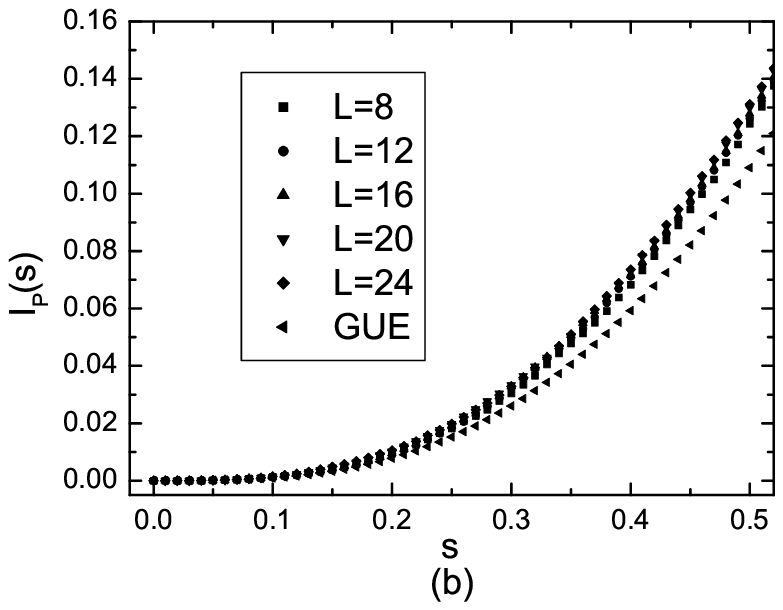}
\end{center}
\begin{center}
 \includegraphics[width=8cm,height=4.9cm]{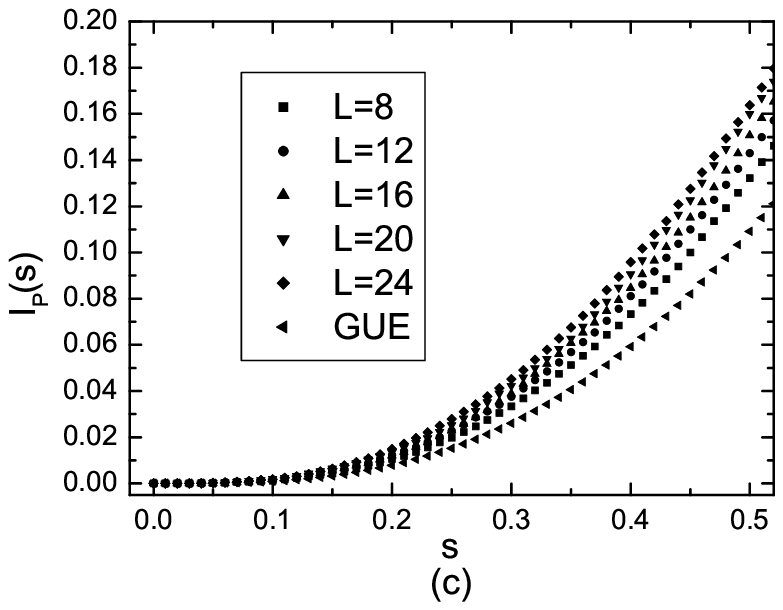}
\end{center}
\caption{$I_P(s)$ vs. $s$ for ${\it L}=8, 12, 16, 20, 24$ and that
for $P_{GUE}(s)$, (a) $E=0$ and $J=0.7$; (b) $E=0.02$ and $J=0.7$;
(c) $E=0.5$ and $J=0.5$.} \label{ps_small2}
\end{figure}
\begin{figure}[ht]
\begin{center}
 \includegraphics[width=8cm,height=4.9cm]{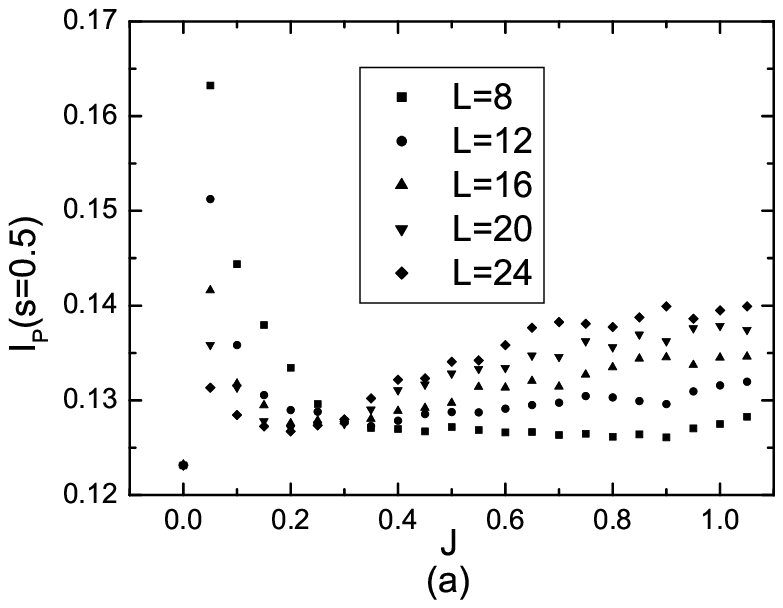}
\end{center}
\begin{center}
 \includegraphics[width=8cm,height=4.9cm]{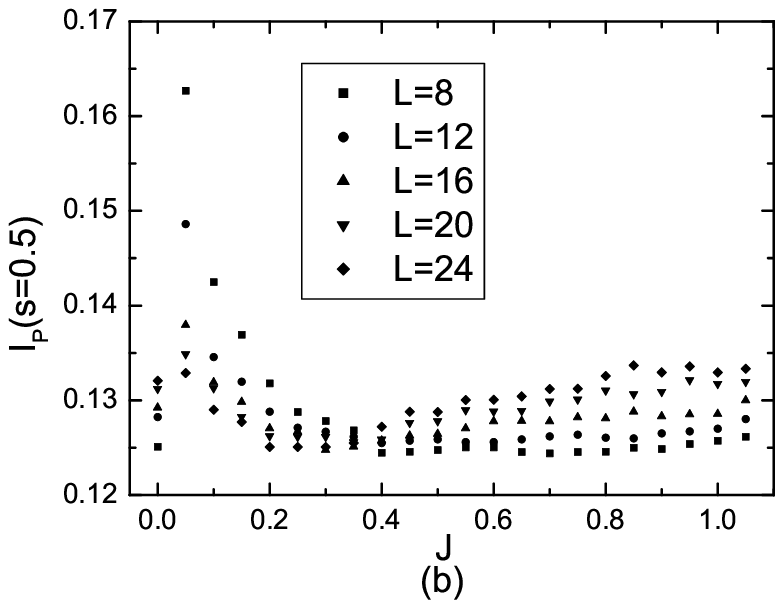}
\end{center}
\begin{center}
 \includegraphics[width=8cm,height=4.9cm]{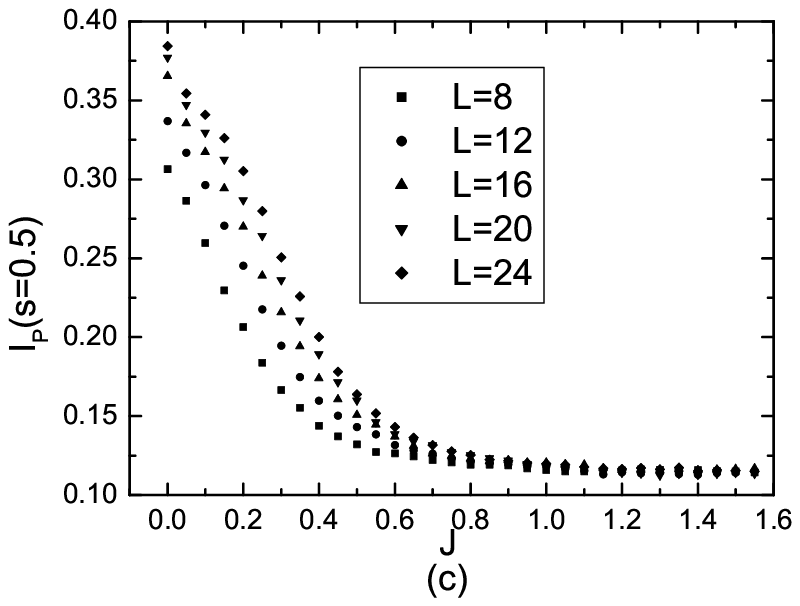}
\end{center}
\caption{$I_P(s=0.5,J)$ vs. $J$ for ${\it L}=8, 12, 16, 20, 24$,
(a) $E=0$; (b) $E=0.02$; (c) $E=0.5$.} \label{ps_small3}
\end{figure}
\begin{figure}[ht]
\begin{center}
 \includegraphics[width=8cm,height=4.9cm]{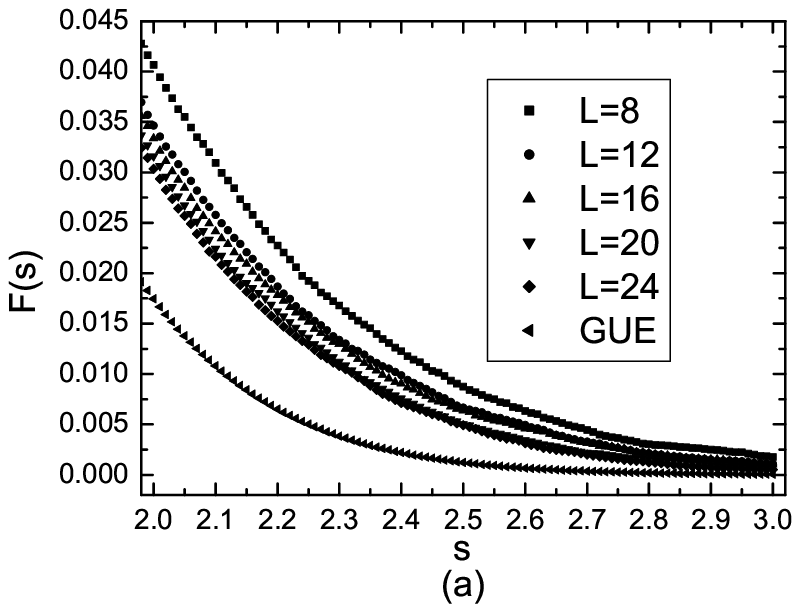}
\end{center}
\begin{center}
 \includegraphics[width=8cm,height=4.9cm]{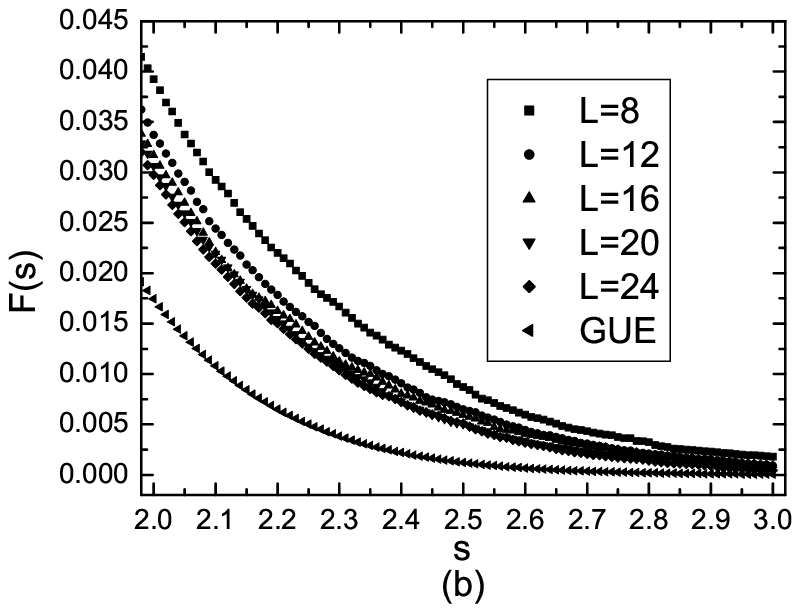}
\end{center}
\begin{center}
 \includegraphics[width=8cm,height=4.9cm]{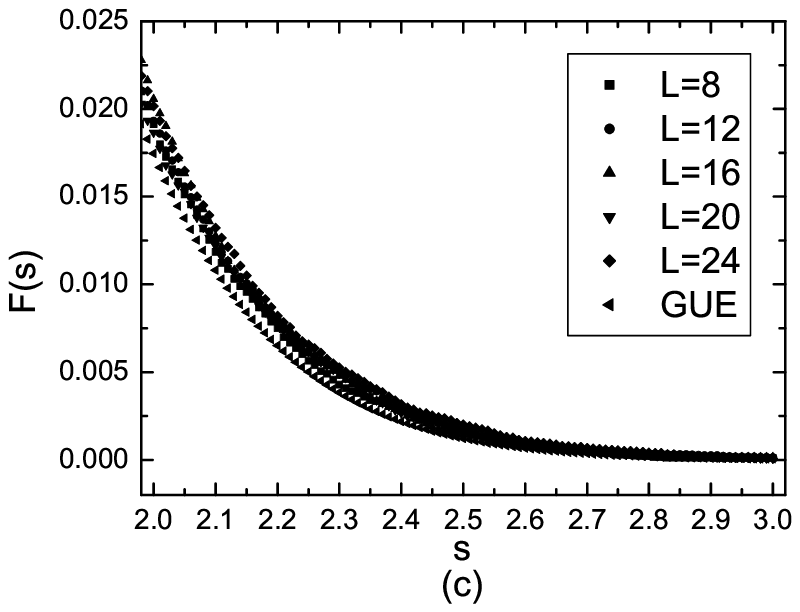}
\end{center}
\caption{$F(s)$ vs. $s$ for ${\it L}=8, 12, 16, 20, 24$, and that
for $P_{GUE}(s)$. (a) $E=0$ and $J=0.1$; (b) $E=0.02$ and $J=0.1$;
(c) $E=0.5$ and $J=1.5$.} \label{ps_tail1}
\end{figure}
\begin{figure}[ht]
\begin{center}
 \includegraphics[width=8cm,height=4.9cm]{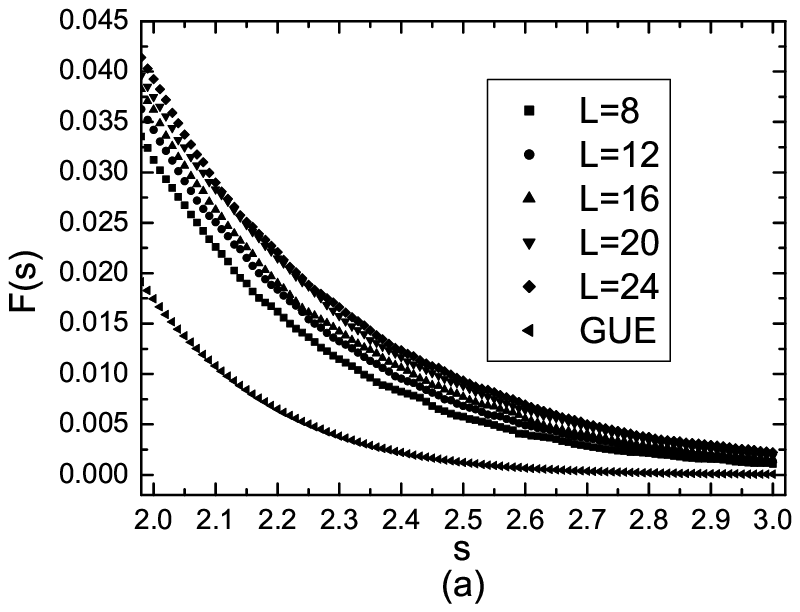}
\end{center}
\begin{center}
 \includegraphics[width=8cm,height=4.9cm]{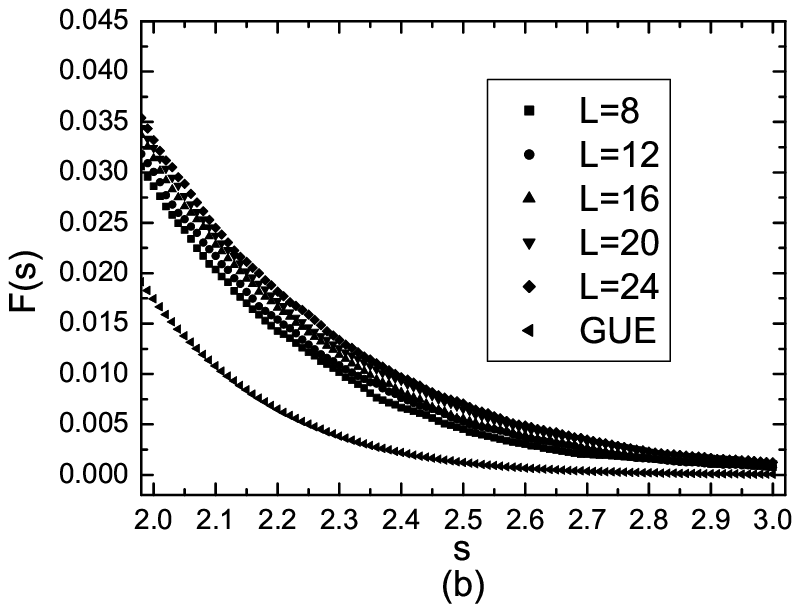}
\end{center}
\begin{center}
 \includegraphics[width=8cm,height=4.9cm]{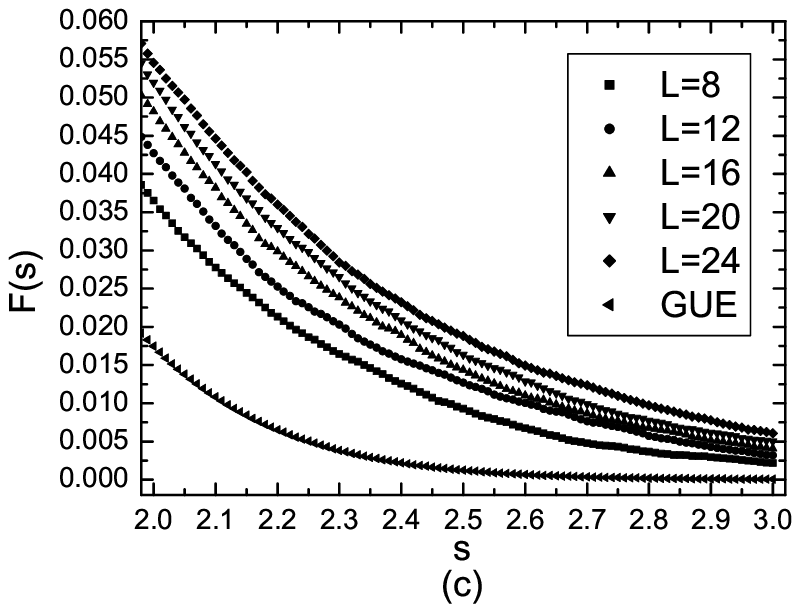}
\end{center}
\caption{$F(s)$ vs. $s$ for ${\it L}=8, 12, 16, 20, 24$ and that
for $P_{GUE}(s)$. (a) $E=0$ and $J=0.7$; (b) $E=0.02$ and $J=0.7$;
(c) $E=0.5$ and $J=0.5$.} \label{ps_tail2}
\end{figure}
\begin{figure}[ht]
\begin{center}
 \includegraphics[width=8cm,height=4.9cm]{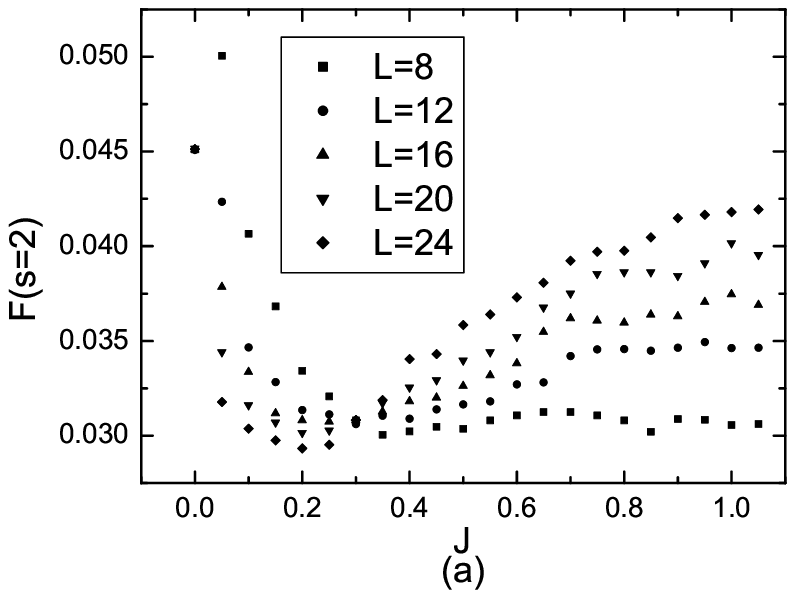}
\end{center}
\begin{center}
 \includegraphics[width=8cm,height=4.9cm]{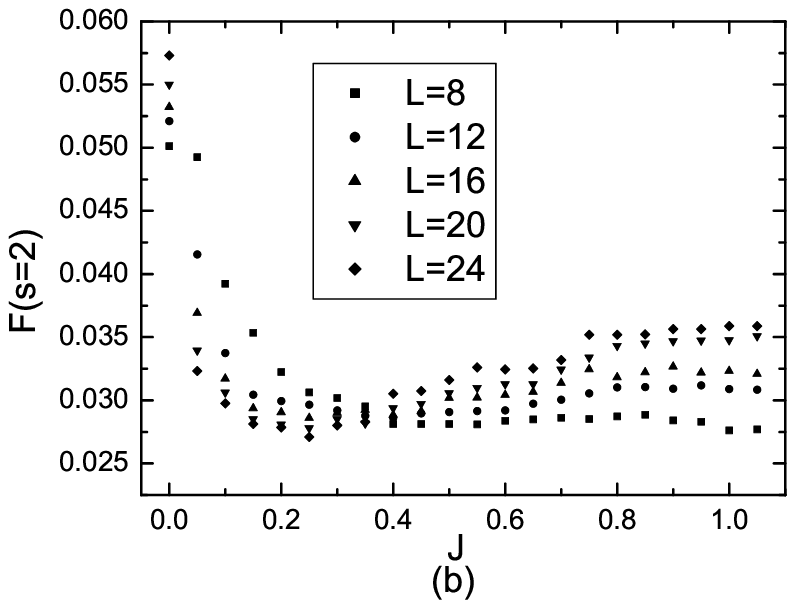}
\end{center}
\begin{center}
 \includegraphics[width=8cm,height=4.9cm]{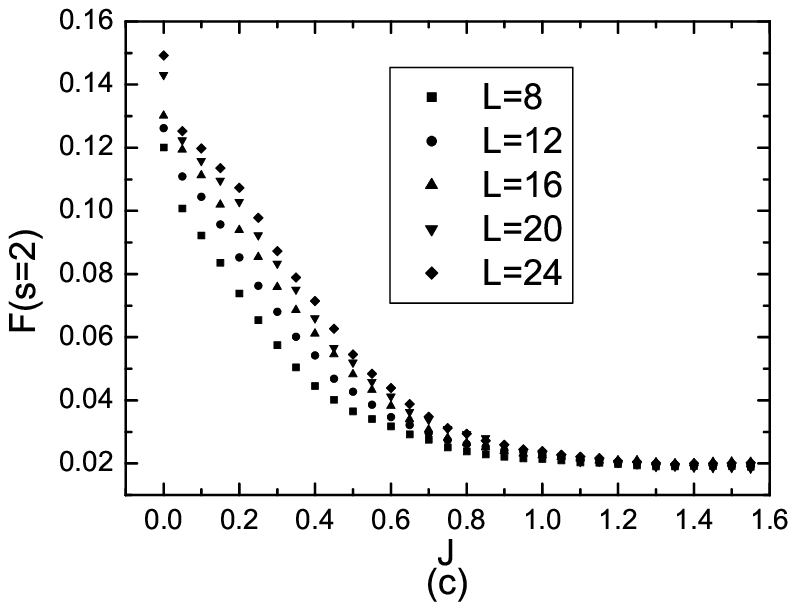}
\end{center}
\caption{$F(s=2,J)$ vs. $J$ for ${\it L}=8, 12, 16, 20, 24$, (a)
$E=0$; (b) $E=0.02$; (c) $E=0.5$.} \label{ps_tail3}
\end{figure}
The overall shape of these curves has some common features. All
curves have a vanishing value when $s$ goes to zero. At small $s$
they increase with $s$ and reach a peak at some intermediate $s$.
Then they decrease monotonically to zero with increasing $s$.
These features are the same as those for
$P_{GUE}(s)$\cite{metzler}. Thus they all look like to be
$P_{GUE}(s)$ at first glance. This raises the question of how to
distinguish numerically extended states from localized states. As
a simple way, it is natural to expect that $P(s)$ for an extended
state approaches $P_{GUE}(s)$ while that for a localized state
should deviate from $P_{GUE}(s)$ as ${\it L}$ increases. Indeed,
as ${\it L}$ increases, curves in each sub-figure of
Fig.\ref{ps_global1} approach $P_{GUE}(s)$ while those in
Fig.\ref{ps_global2} show the opposite tendency. Thus we can use
this different {\it tendency} of $P(s)$ to distinguish extended
states from localized states. We shall show quantitatively that
such opposite tendency for extended states and localized states
exists also in several other characteristic quantities.

Let us first consider a characteristic quantity $I_0$ defined by
$I_0=\int s^2P(s)ds/2$. It is commonly used to characterize the
overall shape of $P(s)$ and to examine the localization
property\cite{metzler}. It is well-known that $I_0=1$ for
localized states while $I_0<1$ for extended states\cite{mehta}.
Thus, the following simple criteria is employed: A state is
localized if its $I_0$ increases and approaches $1$ as ${\it L}$
increases. Otherwise, it is extended. Curves in Fig.\ref{data} are
$I_0$ vs. mixing strength $J$ for $E=0$ (a); 0.02 (b); and 0.5 (c)
for ${\it L}=8, 12, 16, 20, 24$. Fig.\ref{data}(b) shows that the
state of $E=0.02$ is localized at zero mixing and extended at
small $J$. Then it is localized again after $J$ passes a
particular $J_c$ where $I_0$ of different ${\it L}$ cross. For the
state of $E=0$ at the lower band center shown in
Fig.\ref{data}(a), it is extended at zero mixing. Then, it shows
the same feature as the state of $E=0.02$ at small and large $J$.
Fig.\ref{data}(c) shows that state of $E=0.5$ is always localized
at small $J$ and extended only for large $J(>1)$ where all curves
of different system sizes tend to merge together.

A fundamental difference between $P(s)$ for a localized state and
that for an extended states is its behavior at small $s$. As $s$
goes to zero, $P(s)$ vanishes for extended states due to
level-repulsion while it approaches $1$ for localized states due
to level-aggregation\cite{mehta}. Thus we need to consider the
behavior of $P(s)$ at small $s$ for further test of the results in
the last paragraph. It is convenient to consider a function of
integrated level-spacing distribution at small $s$ defined by
$I_P(s)=\int_{0}^{s}P(s^{\prime})ds^{\prime}$. $I_P(s)$ is the
fraction of level-spacing smaller than $s$. Although $P(s)$ in
most cases of our numerical results is close to the GUE
distribution, level-repulsion of extended states and
level-aggregation of localized states should still be expected at
small $s$. This leads to the following criteria: $I_P(s)$ at small
$s$ should increase (decrease) with ${\it L}$ for localized
(extended) states. Thus the behavior of $I_P(s)$ at small $s$ can
serve as another way of distinguishing extended states from
localized ones. Fig.\ref{ps_small1} shows $I_P(s)$ for
$(E=0,J=0.1)$ (a), $(E=0.02,J=0.1)$ (b) and $(E=0.5,J=1.5)$ (c)
for ${\it L}=8, 12, 16, 20,24$ and comparison with $I_P(s)$ of
$P_{GUE}(s)$. Fig.\ref{ps_small2} is for $(E=0,J=0.7)$ (a),
$(E=0.02,J=0.7)$ (b) and $(E=0.5,J=0.5)$ (c). One can see clearly
that states in Fig.\ref{ps_small1} show the feature of extended
states while states in Fig.\ref{ps_small2} are localized. In order
to examine an electronic state of fixed energy in the whole range
of mixing, we consider $I_P(s=0.5)$, the fraction of the
level-spacings less than $0.5$. We plot the results of
$I_P(s=0.5)$ vs. $J$ at $E=0,0.02$ and $0.5$ for {\it L}=8, 12,
16, 20, 24 in Fig.\ref{ps_small3}. Similar to the criteria for
$I_P(s)$, we use the following ones. If $I_P(0.5)$ of a state
increases with ${\it L}$, the state is localized. Otherwise, it is
extended. According to this criteria, curves in
Fig.\ref{ps_small3} give essentially the same results as those
obtained from $I_0$ in Fig.\ref{data}.

Let us now turn to the region of large $s$. Since $P_{GUE}(s)$
decays faster than $P_{PE}(s)$ at large $s$, the behavior of
$P(s)$ in this region can also be used to differentiate extended
and localized states. In this region it is convenient to consider
another quantity defined by $F(s)=\int_{s}^{\infty}
P(s)ds=1-I_P(s)$. The meaning of $F(s)$ is the integrated fraction
of level-spacings larger than $s$. Since $F(s)$ of $P_{GUE}(s)$ is
less than that of $P_{PE}(s)$ at large $s$, we may expect that
$F(s)$ at larger $s$ decreases (increases) with ${\it L}$ for
extended (localized) states. Fig.\ref{ps_tail1} is $F(s)$ for
$P_{GUE}(s)$ and that for $(E=0,J=0.1)$ (a), $(E=0.02,J=0.1)$ (b),
and $(E=0.5,J=1.5)$ (c) with ${\it L}=8, 12, 16, 20, 24$.
Fig.\ref{ps_tail2} is $F(s)$ vs. $s$ for $(E=0, J=0.7)$ (a),
$(E=0.02,J=0.7)$ (b), and $(E=0.5,J=0.5)$ (c). In view of
Fig.\ref{ps_small1} and Fig.\ref{ps_small2}, it is clear that the
results of $F(s)$ coincide with those of $I_P (s)$ concerning the
localization property. We also calculate $F(s=2)$ for fixed energy
states in the whole range of mixing. As shown above, the same
criteria as that for $I_0$ and $I_P(s=0.5)$ can be employed. The
curves of $F(s=2)$ vs. $J$ are plotted in Fig.\ref{ps_tail3} for
$E=0$ (a), $E=0.02$ (b) and $E=0.5$ (c). One can see that they are
consistent with the results of $I_0$ (Fig.\ref{data}) and
$I_P(s=0.5)$ (Fig.\ref{ps_small3}).

\subsection{Discussion of finite-size effect}

In this subsection, we shall consider possible influence of
finite-size effect on our numerical results. It is known that
localization lengths for 2D models can exceed $3\times10^4$
lattice spacings, a thousand times larger than the maximum lattice
size ${\it L}=24$ in our numerical calculation. Therefore, one
should worry about finite-size effect, and question the
unsuitability of our simple criteria for localization property. In
the following discussion, we shall only examine the results of the
quantity $I_0$. Essentially the same discussions can be made for
other quantities such as $I_P(s)$ and $F(s)$.

Let us first consider the results for $E=0$ and $E=0.02$ in
Fig.\ref{data}. The two cases are quite similar. Curves for
different ${\it L}$ cross at a single point $J_c$. As ${\it L}$
increases, $I_0$ decreases and approaches the value for extended
states when $J<J_c$, while it increases and approaches the value
for localized states when $J>J_c$. A straightforward way of
interpreting this behavior is that the state exhibits a transition
from an extended state in $J<J_c$ to a localized one in $J>J_c$.
The correlation length diverges at critical point $J=J_c$ and
decreases sharply when $J$ is slightly away from $J_c$. (In a
metallic phase the correlation length is small while the
localization length is divergent.) Fluctuations at length scale of
order of the correlation length cause $I_0$ to deviate from its
thermodynamic-limit value, but $I_0$ approaches its
thermodynamic-limit values for both $J<J_c$ and $J>J_c$ as lattice
size increases. This is a natural explanation of the results. The
only finite-size effect in this explanation is rounding behavior
in a region close to the critical point $J=J_c$. This region is
normally very narrow because of the sharp drop of the correlation
length near $J=J_c$.

Another possible interpretation is to assume that states are
always localized in the whole region except at $J=J_c$. In this
case, the localization length equals to the correlation length.
Then, in order to explain the above behavior, we have to assume
the following features of the localization length: (a) In the
region $J<J_c$, the ratio of localization length $\xi({\it L})$
and system size ${\it L}$ should increase with ${\it L}$ for small
${\it L}$, leading to a false metallic behavior, while this ratio
decreases with ${\it L}$ for large ${\it L}$, a behavior for the
localized state; (b) in the region $J>J_c$, $\xi({\it L})/{\it L}$
should always decrease with ${\it L}$. In principle, one cannot
rule out this possibility without doing calculations for large
lattice sizes. However, (a) is too strange to justify.
Furthermore, there is no reason to believe why (a) occurs in
region $J<J_c$, but not in region $J>J_c$.

For $E=0.5$ in Fig.\ref{data}, localized behavior is clearly
seen for small and intermediate $J$ while the curves of
different system size tend to merge at large $J$.
Here the finite-size effect should be considered seriously.
There are two ways to explain the merging behavior. One is
that a line of critical points exists for large $J$ where the
correlation length is always divergent. The other is that the
correlation length at the thermodynamic limit is very large
yet {\it finite} and the merging behavior is just a
finite-size effect. Unlike the case of $E=0$ and $E=0.02$,
both explanations in this case are reasonable. The only way
to make an unambiguous conclusion is to do calculations
for large sizes.

However, we can propose a physical picture for the existence of
new extended states at $E\sim0.5$ in the case of strong interband
mixing. Assuming that the intra-band tunnelling at nodes are
negligibly weak for states of $E\sim0.5$, we saw already from Fig.
\ref{network}(a) that the maximum interband mixing
($\sin\theta=1$) delocalizes the state, which is localized at zero
interband mixing. If one views $p=\sin^2 \theta$ as connection
probability of two neighboring loops of opposite chirality, our
two-channel model without intra-band tunnellings at nodes is
analogous to a bond-percolation problem. It is well-known that a
percolation cluster exists at $p\ge p_c=1/2$ or $J\ge J_c=1$ for a
square lattice\cite{stauff}. Therefore, an extended state is
formed by strong mixing. One hopes that the intra-band tunnellings
at nodes will only modify the threshold value of the mixing
strength. If this picture is correct, the finite-size effect only
affects our efforts to determine the accurate value of $J_c$ yet
it does not influence the existence of such a critical point.

It should be noted that all above discussions are based on the
single parameter scaling argument. Suppose that the region of
extended states in thermodynamic limit is vanishing instead of
remaining finite, our above numerical results can also be
explained by introducing irrelevant length scales and considering
their corrections to scaling, as has been pointed out by
Pruisken\cite{pruisken2} and Huckstein\cite{huckstein}. Thus both
our picture of finite region of extended states and the idea of
corrections to scaling are alternative explanations for
non-scaling behaviors. Therefore, calculations for larger sizes of
system are necessary to make unambiguous conclusion for the width
of extended state region in thermodynamic limit. In a recent
theoretical study, Pruisken et. al.\cite{pruisken3} have developed
a microscopic theory based on nonlinear $\sigma$ model to explain
the non-scaling behaviors within the assumption of a single
critical point. However, in their consideration of the case of
long-range correlated disorder which corresponds to the network
model, interband mixing, the physical reason for the existence of
finite region of extended states in our picture, is neglected.
Therefore, their study does not rule out the possibility of finite
region of extended states.

\subsection{Discussion of localization property}

In the past two subsections, we analyzed the global shape of
$P(s)$ and its behavior at small and large $s$ by considering
$I_0$, $I_P(s)$ and $F(s)$, respectively, and discuss possible
influence of finite-size effect. Analysis of all these quantities
leads to essentially the same conclusion concerning the
localization property, as follows. For zero interband mixing, only
states at the two LB centers are extended. In the presence of
interband mixing, new extended states emerge. States near the LB
centers,--- i.e., $E\sim0$,--- are delocalized by weak interband
mixing and localized by strong mixing, with a transition point at
some intermediate mixing $J_c$. For states near the region between
two LBs,---i.e., $E\sim0.5$,--- they are localized at both weak
and intermediate mixing and delocalized by strong mixing.

\begin{figure}[ht]
\begin{center}
 \includegraphics[width=8cm,height=4.9cm]{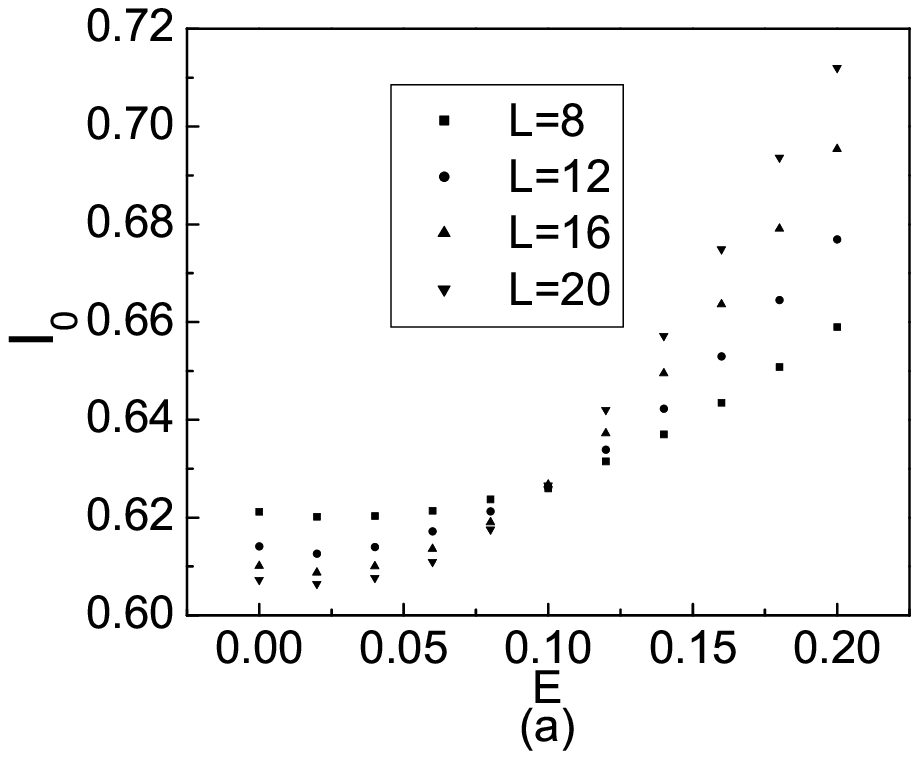}
\end{center}
\begin{center}
 \includegraphics[width=8cm,height=4.9cm]{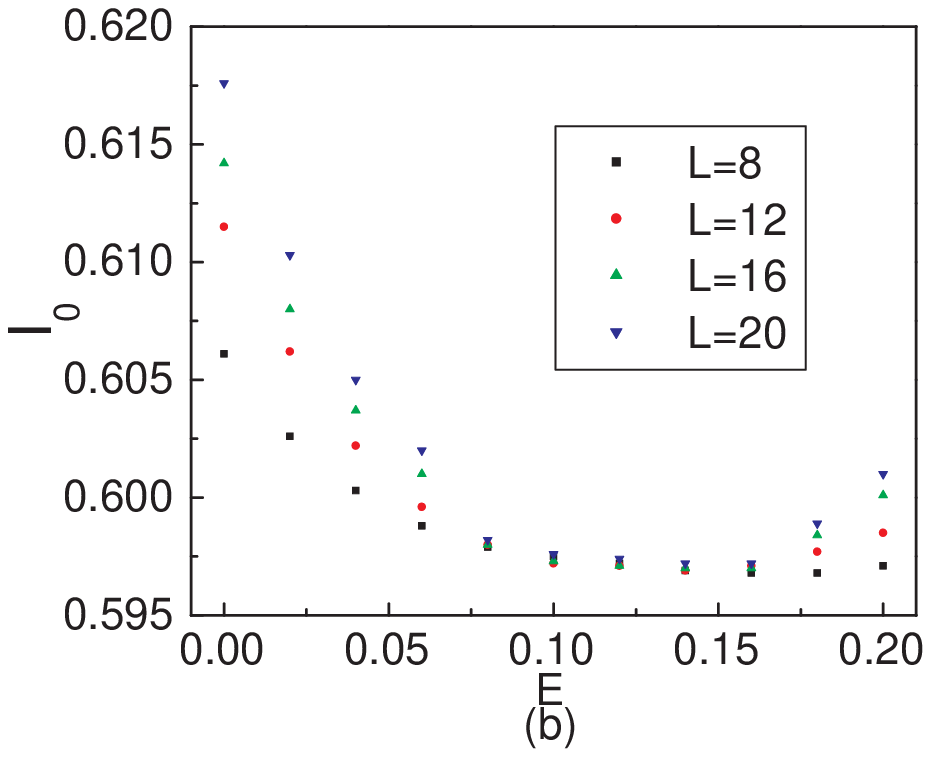}
\end{center}
\begin{center}
 \includegraphics[width=8cm,height=4.9cm]{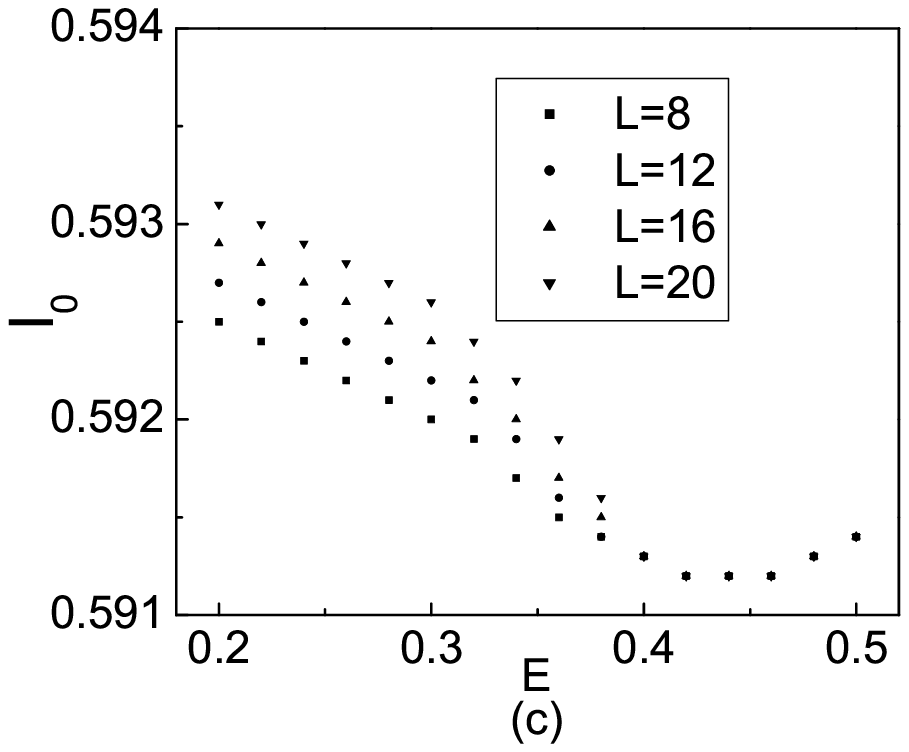}
\end{center}
\caption{$I_0$ vs. $E$ for ${\it L}=8, 12, 16, 20$, (a) $J=0.1$;
(b) $J=0.7$; (c) $J=1.5$.} \label{fixed_J}
\end{figure}

In order to show explicitly the existence of a narrow band of
extended states and its evolution with increasing mixing, curves
of $I_0$ vs. $E$ are plotted for three values of $J$ in
Fig.\ref{fixed_J}. A band of extended states is formed around the
LB center $E\in[0,0.1]$ for $J=0.1$. When $J$ is increased to an
intermediate value $0.7$, this band of extended states is lifted
up to $E\in[0.8,1.6]$. For strong mixing, it is further shifted to
$E\in[0.4,0.5]$. Thus the band of extended states in the lower LB
tends to float up in energy while that of the upper tends to dive
down in energy as mixing strength increases. The two bands should
finally meet at the middle energy region in the case of strong
mixing.

The above results are restricted to the case of two LBs. However,
there are infinite number of LBs in a realistic system. In order
to conjecture the situation when infinite number of LBs is
incorporated in our result, we take into account the
float-up-merge picture proposed by Sheng et. al.\cite{sheng2}. We
shall expect that a narrow extended band appears in each LB center
for weak interband mixing. Increasing mixing, i.e., increasing
disorders or decreasing magnetic field, the extended band in the
lowest LB floats up and finally merges with that in the second
lowest LB. Then, this extended band will further shift up and
merge into that in the third lowest LB, and so on so forth.

\begin{figure}[ht]
\begin{center}
 \includegraphics[width=4cm,height=4cm]{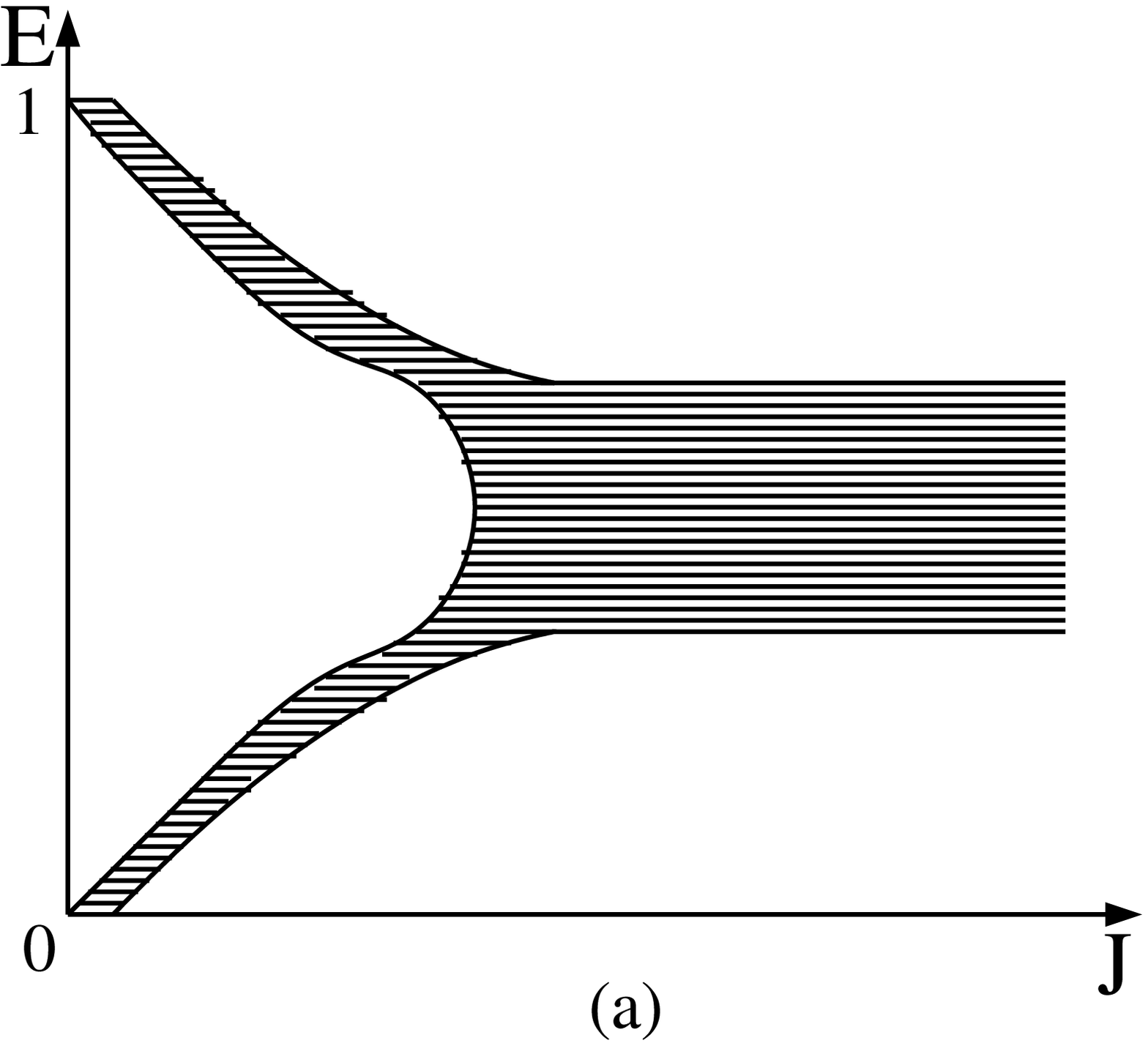}
 \includegraphics[width=4cm,height=4cm]{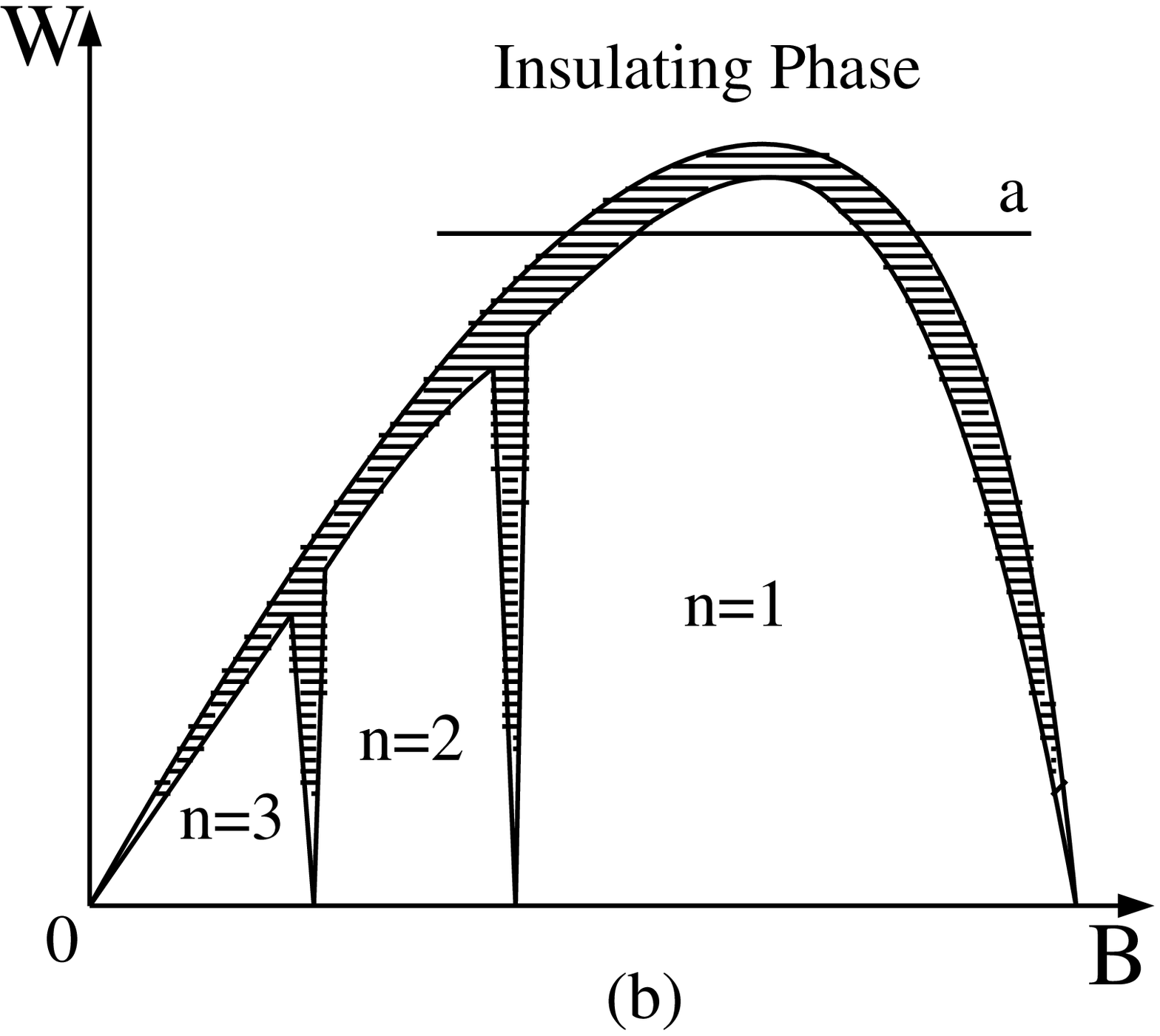}
\end{center}
\caption{(a) Topological phase diagram of electron localization in
$E-J$ plane. The shadowed regime is for extended states (metallic
phase). (b) Topological QH phase diagram in $W-B$ plane. $W$
stands for the disorder strength, and $B$ for the magnetic field.
The shadowed regime is for the metallic phase. The area indicated
by the symbol $n$ is the $n$-plateau IQHE phase. The rest area is
the insulating phase.} \label{phase}
\end{figure}

To express our numerical results in the plane of energy and
interband mixing, a topological phase diagram shown in
Fig.\ref{phase}(a) is obtained. In the absence of interband
mixing, only the singular energy level at each LB center is
extended. In the presence of interband mixing of opposite
chirality, there are two regimes. At weak mixing, each of the
extended states broadens into a narrow band of extended states
near the LB centers. With increased mixing, the extended states in
the lowest LB shift from the LB center(see Fig.\ref{fixed_J}).
These extended states will eventually merge with those from the
higher LBs. This shifting of bands of extended states is similar
to the shifting of single extended states at LB centers observed
in previous studies \cite{kivelson} where emerging of extended
bands is missing. At strong mixing, a band of extended states
exists between neighboring LBs where all states are localized
without the mixing.

Let us look at the consequences of the above results. For weakly
disordered systems in IQHE regime, the Landau gap is larger than
the LB bandwidth. Thus there is no overlap between adjacent LBs.
According to the semiclassical picture, electronic states between
the two adjacent LBs should be from either the upper or the lower
bands with the same chirality in this case. It means that no
interband mixing occurs and there is only one extended state in
each LB. This may explain why scaling behaviors were observed for
plateau transitions in early experiments on clean samples.
Interband mixing occurs when the Landau gap is less than the LB
bandwidth. Systems of relatively strong disorders in IQHE regime
should correspond to this case. As the single extended state at
each LB center broadens into a narrow extended band, a narrow
metallic phase emerges between two neighboring IQHE phases. Thus
each plateau transition contains two consecutive quantum phase
transitions for strongly disordered systems. The bands of extended
states will merge together in strong mixing. This strong mixing
regime corresponds to the case when the Landau gap is much smaller
than the bandwidth. Since the Landau gap is proportional to the
magnetic field, the disordered system should always enter the
strong mixing regime before it reaches the weak field insulating
phase, regardless of how weak the disorders are. In terms of QH
plateau transitions, a direct transition occurs because a narrow
metallic phase exists between two QH phases in a weak field. Thus,
we propose that a direct transition from an IQHE phase to the
insulating phase at weak field is realized by passing through a
metallic phase, and it should hold for both weak and strong
disordered systems.

Plotting above results in the plane of disorder and the magnetic
field, we obtain a new topological QH phase diagram as shown in
Fig.\ref{phase}(b). This is similar to the empirical diagram
obtained experimentally in Ref. 15. The origin ($W=0, \ B=0$) is a
singular point. According to the weak localization
theory\cite{abrahams}, no extended state exists at this point.
Differing from existing theories, there exists a narrow metallic
phase between two adjacent IQHE phases and between an IQHE phase
and an insulating phase.

\subsection{Comparison with previous studies}
\begin{figure}
\begin{center}
 \includegraphics[width=8cm,height=5cm]{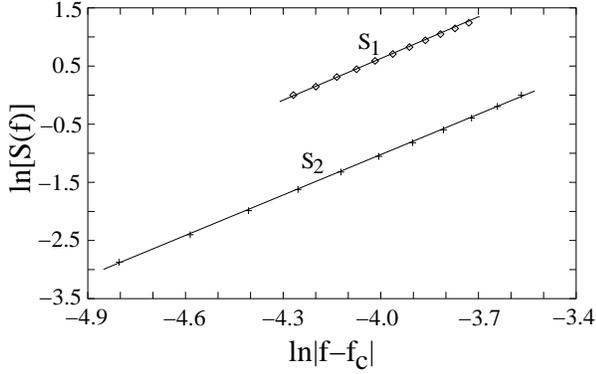}
\end{center}
\caption{The fitting result of two critical points at the left and
the right side. The two straight lines show coincidence with the
scaling law. The critical filling factors are $f_{c1}=0.6453$ and
$f_{c2}=0.6477$. The two critical exponents are equal to the value
$z\nu=2.33\pm0.01$.} \label{twopoint}
\end{figure}

\begin{figure}
\begin{center}
 \includegraphics[width=8cm,height=5cm]{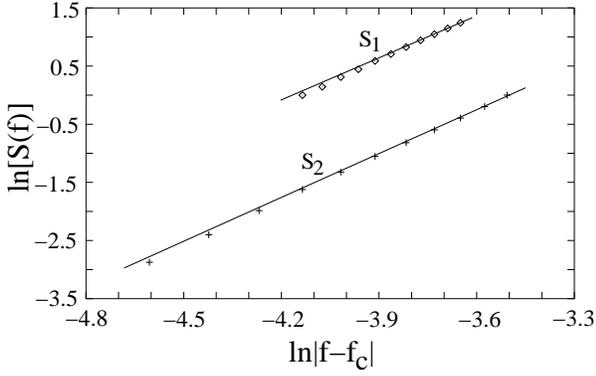}
\end{center}
\caption{The best fitting result of one single critical point at
the left and right side. The critical filling factor is
$f_{c}=0.646$. The two straight lines illustrate systematic
deviations from the scaling law at regions close to the critical
point. The average values of the two critical exponents are
$z_1\nu_1=2.58\pm0.02$ and $z_2\nu_2=2.60\pm0.02$, respectively.}
\label{onepoint}
\end{figure}

In this subsection, we shall compare our results with previous
studies. We shall show that the new phase diagram in
Fig.\ref{phase} is consistent with the non-scaling
experiments\cite{hilke} where samples are relatively dirty, and
interband mixing is strong, corresponding to a process along line
$a$ in Fig.\ref{phase}(b). The system undergoes two quantum phase
transitions each time when it moves from the QH insulating phase
to IQHE phase of $n=1$ and back to the weak field insulating phase
as the magnetic field decreases.
\begin{figure}[ht]
\begin{center}
\includegraphics[width=8cm,height=5cm]{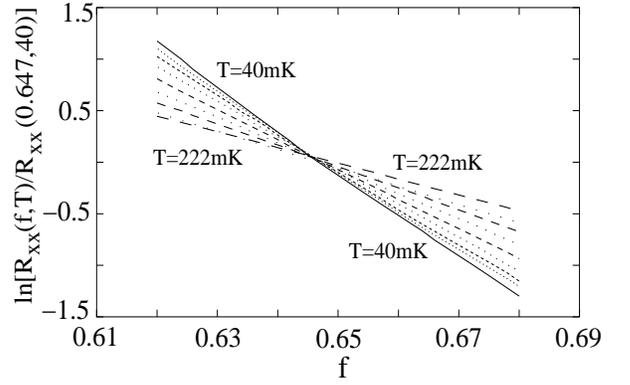}
\end{center}
\caption{\label{experiment} The original experimental data of
$\ln[R_{xx}(f,T)/R_{xx}(0.647,40mK)]$ in Ref. 18.
$f$ is the filling factor of LBs and $T$ is the temperature.}
\end{figure}
To verify this claim, we analyzed the original experimental data in
Ref. 18 according to the assumption of two quantum 
phase transition points. The experiment data of the logarithm of
the longitudinal resistance $ln[R_{xx}(f,T)]$ are shown in
Fig.\ref{experiment} where $f$ is the filling factor of LBs and
$T$ is the temperature. According to the theory of continuous
transitions, one should obtain
\begin{equation}
    ln[R_{xx}(\nu,T)]=F_1(S_1(f)/T)
\end{equation}
with $S_1(f)\sim(f_{c1}-f)^{z_1\nu_1}$ for the region of
$f<f_{c1}$ while
\begin{equation}
    ln[R_{xx}(\nu,T)]=F_2(S_2(f)/T)
\end{equation}
with $S_2(f)\sim(f-f_{c2})^{z_2\nu_2}$ for the region of
$f>f_{c2}$. Previous theories predict one single critical point,
---i.e., $f_{c1}=f_{c2}$ and $z_1\nu_1=z_2\nu_2$. But our results
suggest two distinct critical points. By standard scaling
analysis, two good scaling behaviors are obtained for two close
critical filling factors of $f_{c1}=0.6453$ and $f_{c2}=0.6477$ as
shown in Fig.\ref{twopoint}. The critical exponents in both the
left side and the right side of the transition region are equal to
the value $z\nu=2.33\pm0.01$. On the other hand, the fit for one
single critical point fails. Fig.\ref{onepoint} shows the result
of a single critical point at $\nu_c=0.646$. It is the best
fitting result for a single critical point if we require that the
two critical exponents are approximately equal and the scaling law
is optimally obeyed. The two critical exponents are
$z_1\nu_1=2.58\pm0.02$ and $z_2 \nu_2 = 2.60 \pm 0.02$, deviating
from the theoretical results $z\nu \sim 2.33$. One can also see
clearly systematic deviations from the scaling law in the region
close to the critical point at both sides in Fig.\ref{onepoint}.
This implies that the transition process is governed by two
separated critical points instead of one. The regime between the
two critical points should correspond to the metallic phase.

It is worth noting that there is another puzzle in the non-scaling
experiment which may be solved by our two-critical-point picture.
As an example, we consider the experimental data for the transition
between the QH insulating phase and the $n=1$ IQHE phase.
It was shown\cite{hilke} that the logarithm of the longitudinal
resistance $ln[R_{xx}(f,T)]$ can be fitted by a linear function
of the filling factor $f$ (see Fig.\ref{experiment})
\begin{equation}
  ln[R_{xx}(f,T)]=ln[R_{xx}(f_c,T)]-(f-f_c)/(\alpha+\beta T)]
\end{equation}
where $\alpha$ and $\beta$ are positive constants, $f_c$ is the
filling factor where curves of different temperature $T$ cross
approximately. Since $\alpha$ is non-zero\cite{hilke}, it leads to
the conclusion that $R_{xx}(f,T)$ at the limit of $T=0$ remains
finite {\it for every $f$}. This is puzzling because it is
inconsistent with the theoretical requirement that
$R_{xx}(T=0)=\infty$ in the QH insulating phase, i.e., $f<f_c$,
and $R_{xx}(T=0)=0$ in the $n=1$ IQHE phase, i.e., $f>f_c$. This
puzzle may be solved as follows. Combine the linear relationship
between $ln[R_{xx}(f,T)]$ and $f$ for fixed $T$ with our picture
of two critical points $f_{c1}<f_{c2}$, we expect
\begin{equation}
  ln[R_{xx}(f,T)]=ln[R_{xx}(f_{c1},T)]-(f-f_{c1})/(A_1 T^{z\nu})
\end{equation}
in the QH insulating phase, i.e., $f<f_{c1}$, while
\begin{equation}
  ln[R_{xx}(f,T)]=ln[R_{xx}(f_{c2},T)]-(f-f_{c2})/(A_2 T^{z\nu})
\end{equation}
in the n=1 IQHE phase, i.e., $f>f_{c2}$, where $A_1$ and $A_2$ are
positive constants, and $z$ and $\nu$ are critical exponents. It
is clear that both $R_{xx}(f,T=0)=\infty$ in $f<f_{c1}$ and
$R_{xx}(f,T=0)=0$ in $f>f_{c2}$ are recovered. While a finite
value of $R_{xx}(f,T=0)$ in the region $f_{c1}<f<f_{c2}$ is
consistent with our prediction of a metallic phase between the two
critical points.

Scaling behaviors of plateau-plateau transitions are observed in
recent experiments\cite{hohls1,hohls2,hohls3,ponomarenko,li}. At a
first glance, it seems that these results conflict with both
non-scaling experiments and our numerical results. However, there
are two important differences between recent scaling experiments
and non-scaling experiments. One is that some scaling experiments
\cite{hohls1,hohls2,hohls3} were done deep inside QH plateau, or
far from the plateau transition point, while the non-scaling
behavior was obtained by using data very close to the transition
points. In the non-scaling experiments\cite{hilke}, it is known
that data not too close to a transition point follow a scaling
law. The other is that the samples used in all scaling
experiments\cite{hohls1,hohls2,hohls3,ponomarenko,li} are clean
with very high mobility while non-scaling behavior was observed in
relative dirty samples\cite{hilke,shash,baba}. In fact, the
mobility in recent scaling experiments is at least one order of
magnitude larger than that in early non-scaling experiments. This
means that the scaling and non-scaling experiments correspond to
regions of vanishing and relatively strong inter-band mixing,
respectively. Thus, there is no real conflict between the recent
scaling experiments and those early non-scaling experiments. Since
our model is valid only when inter-band mixing is considerable,
there is also no real conflict between recent scaling experiments
and our numerical results.

Two-channel CC model has been used to simulate two degenerated or
nearly-degenerated spin-resolved Landau subbands with strong
interband mixing by Wang {\it et al.}\cite{wang}. They found
two distinct critical points which were related to mixing-induced
repulsion\cite{wang}. For the nearly-degenerated case, they did
not consider the states between the two LB centers. For the
degenerated case, their study could not discern whether the states
between the two critical points are extended or localized.
According to our results, a band of extended states is formed
around the degenerated LB center for the degenerated case at
strong mixing, while for the nearly-degenerated case electronic
states between the two LB centers are delocalized by strong mixing.
Thus our results are consistent with their results.

One should also notice that two types of metallic phases have been
studied extensively in the QH system. One is the composite Fermion
state at the half-filling in the lowest Landau level (LL) and the
other is the stripe state at the half-filled higher LLs. These
states are formed by the Coulomb interaction effect in the high
mobility samples. They are different from our metallic phase
caused by level mixing. Although we have not considered the
electron-electron interactions in our study, there is no reason
why the delocalization effect of level mixing will be diminished
by the Coulomb interaction. Of course, the interaction could
lead to a level shift, thus it may modify the band width.

\section{Conclusions}
In conclusion, we find by numerical calculations within the
network model that it is possible that the single extended state
at each LB center in the absence of interband mixing may broaden
into a narrow band of extended states when the effect of mixing of
states of {\it opposite chirality} is taken into account. We also
provide a physical picture to show how the mixing of states of
{\it opposite chirality} may possibly lead to the existence of
extended state bands. Based on above results, we propose a new
phase diagram in which a narrow metallic phase exists between two
neighboring IQHE phases and between an IQHE phase and an
insulating phase. This new phase diagram is consistent with
non-scaling behaviors observed in recent experiments. A standard
scaling analysis on nonscaling experiment data \cite{hilke}
supports our results. However, due to finite-size effects, our
numerical results can also be explained based on the assumption of
a single critical point. Thus further study on large system size
is still needed to conclude whether there are extended state bands
in quantum Hall systems in thermodynamic limit.

\begin{acknowledgments}
This work was substantially supported by a grant from the Research
Grant Council of HKSAR, China (Project No. HKUST6153/99P, and
HKUST6149/00P). GX acknowledges the support of CNSF under grant
No. 10347101 and the Grant from Beijing Normal University. GX and
YW also acknowledge the support of CNSF under grant No. 90103024.
\end{acknowledgments}

\appendix*
\section{}
\begin{figure}[ht]
\begin{center}
 \includegraphics[height=7cm, width=8cm]{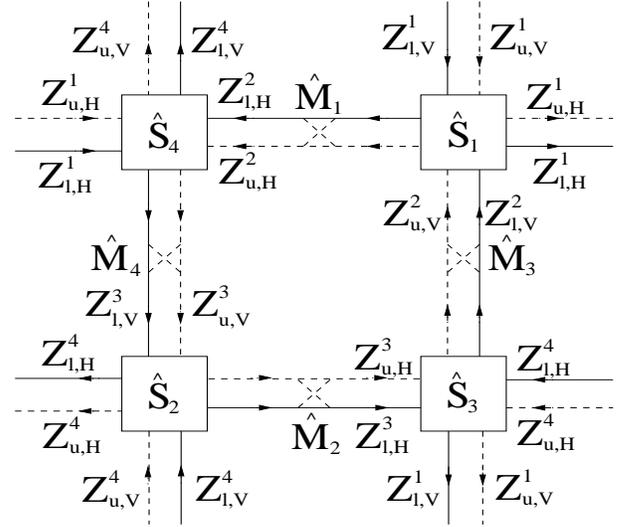}
\end{center}
\caption{A two-channel network model of $2\times2$ nodes with
periodic boundaries along both directions. $Z$-s are the
wavefunction amplitudes on links. The notations are as follows. H
and V stand  for horizontal and vertical links, respectively. $u$
($l$) is for the upper (lower) LB. $\hat{S}_i$ are SO(4) matrices
describing tunnelling at nodes, and $\hat{M}_i$ are $U(2)$
matrices for interband mixing.} \label{example}
\end{figure}

In this appendix, we explicitly construct the evolution matrix
$\hat U$ for a $2\times 2$ two-channel CC-network model as shown
in Fig.\ref{example}. Periodical boundary conditions in both
directions are imposed as explained in section III. $Z$-s are the
wavefunction amplitudes on links. The notations are as follows. H
and V stand  for horizontal and vertical links, respectively. $u$
($l$) is for the upper (lower) LB. $\hat{S}_i$ are SO(4) matrices
defined in Eq.(\ref{smatrix}) describing the tunnelling on nodes,
and $\hat{M}_i$ are $U(2)$ matrices defined in Eqs. \ref{mmatrix1}
and \ref{mmatrix2} describing interband mixing. From
Fig.\ref{example} we can obtain
\begin{equation}
\left (
\begin{array}{l}
Z_{u,H}^{1}(t+1) \\ Z_{l,H}^{1}(t+1) \\
Z_{l,H}^{2}(t+1) \\ Z_{u,H}^{2}(t+1)
\end{array}
\right ) =
\hat{H}_1
\left (
\begin{array}{l}
Z_{l,V}^{1}(t)\\ Z_{u,V}^{1}(t)\\ Z_{u,V}^{2}(t) \\
Z_{l,V}^{2}(t)
\end{array}
\right )
\end{equation}
\begin{equation}
\left (
\begin{array}{l}
Z_{u,H}^{3}(t+1) \\ Z_{l,H}^{3}(t+1) \\
Z_{l,H}^{4}(t+1) \\ Z_{u,H}^{4}(t+1)
\end{array}
\right ) =
\hat{H}_2
\left (
\begin{array}{l}
Z_{l,V}^{3}(t) \\ Z_{u,V}^{3}(t) \\ Z_{u,V}^{4}(t)
\\ Z_{l,V}^{4}(t)
\end{array}
\right )
\end{equation}
\begin{equation}
\left (
\begin{array}{l}
Z_{l,V}^{1}(t+1) \\ Z_{u,V}^{1}(t+1) \\
Z_{u,V}^{2}(t+1) \\ Z_{l,V}^{2}(t+1)
\end{array}
\right ) =
\hat{H}_3
\left (
\begin{array}{l}
Z_{u,H}^{3}(t) \\ Z_{l,H}^{3}(t) \\ Z_{l,H}^{4}(t)
\\ Z_{u,H}^{4}(t)
\end{array}
\right )
\end{equation}
\begin{equation}
\left (
\begin{array}{l}
Z_{l,V}^{3}(t+1) \\ Z_{u,V}^{3}(t+1) \\
Z_{u,V}^{4}(t+1) \\ Z_{l,V}^{4}(t+1)
\end{array}
\right ) =
\hat{H}_4
\left (
\begin{array}{l}
Z_{u,H}^{1}(t) \\ Z_{v,H}^{1}(t) \\ Z_{l,H}^{2}(t)
\\ Z_{u,H}^{2}(t)
\end{array}
\right ),
\end{equation}
with
\begin{equation}
   \hat{H}_1=
\left (
\begin{array}{ll}
\hat{1} & \hat{0} \\
\hat{0} & \hat{M}_1
\end{array}
\right )
\hat{S}_1; \quad
   \hat{H}_2=
\left (
\begin{array}{ll}
\hat{M}_2 & \hat{0} \\
\hat{0} & \hat{1}
\end{array}
\right )
\hat{S}_2; \nonumber
\end{equation}
\begin{equation}
   \hat{H}_3=
\left (
\begin{array}{ll}
\hat{1} & \hat{0} \\
\hat{0} & \hat{M}_3
\end{array}
\right )
\hat{S}_3; \quad
   \hat{H}_4=
\left (
\begin{array}{ll}
\hat{M}_4 & \hat{0} \\
\hat{0} & \hat{1}
\end{array}
\right )
\hat{S}_4,  \nonumber
\end{equation}
where $\hat{1}$ and $\hat{0}$ are the $2\times 2$ identity
and zero matrices, respectively.
If we define
\begin{eqnarray}
\phi_{H} = \left (
        \begin{array}{l}
Z_{u,H}^{1}\\ Z_{l,H}^{1}\\
Z_{l,H}^{2}\\ Z_{u,H}^{2}\\
Z_{u,H}^{3}\\ Z_{l,H}^{3}\\
Z_{l,H}^{4}\\ Z_{u,H}^{4}\\
        \end{array}
        \right ); \quad
\phi_{V} = \left (
        \begin{array}{l}
Z_{l,V}^{1}\\ Z_{u,V}^{1}\\
Z_{u,V}^{2}\\ Z_{l,V}^{2}\\
Z_{l,V}^{3}\\ Z_{u,V}^{3} \\
Z_{u,V}^{4}\\ Z_{l,V}^{4} \\
        \end{array}
        \right ), \nonumber
\end{eqnarray}
then the evolution equation is
\begin{equation}
        \left (
        \begin{array}{l}
        \phi_H(t+1) \\ \phi_V(t+1)
        \end{array}
        \right )
        =
        \hat{U}
        \left (
        \begin{array}{l}
        \phi_H(t) \\ \phi_V(t)
        \end{array}
        \right ).
\end{equation}
The evolution operator $\hat{U}$ is
\begin{equation}
        \hat{U}
        =
        \left (
        \begin{array}{llll}
        \hat{0} & \hat{0} & \hat{0} & \hat{H}_1 \\
        \hat{0} & \hat{0} & \hat{H}_2 & \hat{0} \\
        \hat{H}_3 & \hat{0} & \hat{0} & \hat{0} \\
        \hat{0} & \hat{H}_4 & \hat{0} & \hat{0}
        \end{array}
        \right ),
\end{equation}
where $\hat{0}$ is the $4\times 4$ zero matrix. It has the
structure of Eq.(\ref{evolution}).

\end{document}